\documentclass[letterpaper,twocolumn,prL,aps,superscriptaddress,floatfix]{revtex4}

\usepackage[latin9]{inputenc}
\usepackage{amsmath}
\usepackage{amssymb}
\usepackage{graphicx}
\usepackage{esint}
\usepackage{times}

\usepackage[unicode=true,
bookmarks=true,bookmarksnumbered=false,bookmarksopen=false,
breaklinks=false,pdfborder={0 0 1},backref=false,colorlinks=true]
{hyperref}
\hypersetup{
    linkcolor=magenta, urlcolor=blue, citecolor=blue, pdfstartview={FitH}, hyperfootnotes=false, unicode=true}

\makeatletter

\pdfpageheight\paperheight
\pdfpagewidth\paperwidth

\@ifundefined{textcolor}{}
{%
    \definecolor{BLACK}{gray}{0}
    \definecolor{WHITE}{gray}{1}
    \definecolor{RED}{rgb}{1,0,0}
    \definecolor{GREEN}{rgb}{0,1,0}
    \definecolor{BLUE}{rgb}{0,0,1}
    \definecolor{CYAN}{cmyk}{1,0,0,0}
    \definecolor{MAGENTA}{cmyk}{0,1,0,0}
    \definecolor{YELLOW}{cmyk}{0,0,1,0}
}

\usepackage{xcolor}
\usepackage{soul}

\definecolor{blue}{rgb}{0,0,1}
\definecolor{red}{rgb}{1,0,0}
\definecolor{green}{rgb}{0,1,0}

\makeatother

\begin{document}
\title{Quantum squeezing amplification with a weak Kerr nonlinear oscillator}

\author{Yanyan Cai}
\thanks{These authors contributed equally to this work.}
\affiliation{Southern University of Science and Technology, Shenzhen 518055, China}
\affiliation{International Quantum Academy, Shenzhen 518048, China}

\author{Xiaowei Deng}
\thanks{These authors contributed equally to this work.}
\affiliation{International Quantum Academy, Shenzhen 518048, China}

\author{Libo Zhang}
\thanks{These authors contributed equally to this work.}
\affiliation{Southern University of Science and Technology, Shenzhen 518055, China}
\affiliation{International Quantum Academy, Shenzhen 518048, China}

\author{Zhongchu Ni}
\affiliation{International Quantum Academy, Shenzhen 518048, China}

\author{Jiasheng Mai}
\affiliation{Southern University of Science and Technology, Shenzhen 518055, China}
\affiliation{International Quantum Academy, Shenzhen 518048, China}

\author{Peihao Huang}
\affiliation{International Quantum Academy, Shenzhen 518048, China}

\author{Pan Zheng}
\affiliation{International Quantum Academy, Shenzhen 518048, China}

\author{Ling Hu}
\affiliation{International Quantum Academy, Shenzhen 518048, China}
\affiliation{Shenzhen Branch, Hefei National Laboratory, Shenzhen 518048, China}
\author{Song Liu}
\affiliation{International Quantum Academy, Shenzhen 518048, China}
\affiliation{Shenzhen Branch, Hefei National Laboratory, Shenzhen 518048, China}

\author{Yuan Xu}
\email{xuyuan@iqasz.cn}
\affiliation{International Quantum Academy, Shenzhen 518048, China}
\affiliation{Shenzhen Branch, Hefei National Laboratory, Shenzhen 518048, China}

\author{Dapeng Yu}
\affiliation{International Quantum Academy, Shenzhen 518048, China}
\affiliation{Shenzhen Branch, Hefei National Laboratory, Shenzhen 518048, China}

\begin{abstract}
Quantum squeezed states, with biased quantum noise, have been widely utilized in quantum sensing and quantum error correction applications. However, generating and manipulating these nonclassical states with a large squeezing degree typically requires strong nonlinearity, which inevitably induces additional decoherence that diminishes the overall performance. Here, we demonstrate the generation and amplification of squeezed states in a superconducting microwave cavity with weak Kerr nonlinearity. By subtly engineering an off-resonant microwave drive, we observe cyclic dynamics of the quantum squeezing evolution in a displaced frame of the cavity. Furthermore, we deterministically realize quantum squeezing amplification by alternately displacing the Kerr oscillator using the Trotterization technique, achieving a maximum squeezing degree of 14.6 dB and a squeezing rate of 0.28 MHz. Our demonstrated displacement-enhanced squeezing operation offers a hardware-efficient approach for generating large squeezed states, promising potential applications in quantum-enhanced sensing and quantum information processing.
\end{abstract}
\maketitle
\vskip 0.5cm

\noindent \textbf{\large{}Introduction}{\large\par}
\noindent The landscape of quantum information science has been substantially enriched by the utilization of nonclassical bosonic states, which are essential for continuous variable quantum information processing~\cite{braunstein2005,andersen2010}. Among these states, squeezed states~\cite{yuen1976}, which exhibit reduced noise in one quadrature but increased noise in the conjugate one, stand out for their remarkable ability to enhance the measurement sensitivity beyond the standard quantum limit~\cite{caves1981,walls1983}. This enhancement is particularly critical in sensing applications~\cite{schnabel2017,lawrie2019}, such as the search for elusive dark matter particles~\cite{backes2021} and the observation of gravitational waves~\cite{abadie2011,tse2019}. In addition, these nonclassical squeezed states serve as essential resources to generate Gottesman-Kitaev-Preskill states~\cite{fluhmann2019,hastrup2021}, squeezed cat states~\cite{lo2015,pan2023}, and squeezed Fock states~\cite{kienzler2017} for quantum error correction protocols~\cite{campagne2020,schlegel2022,xu2023,korolev2024}.

Over the past few decades, the generation and manipulation of squeezed states have been successfully demonstrated on a variety of physical platforms~\cite{andersen2016}, including optical and microwave photons~\cite{yurke1988,vahlbruch2008,safavi2013,wang2017,dassonneville2021,eickbusch2022,qiu2023,eriksson2024}, as well as mechanical and acoustic phonons~\cite{wollman2015,ma2021,youssefi2023,marti2024,matsos2024}. The underlying mechanism for generating these nonclassical states generally involves engineering a nonlinear interaction within these bosons. The pursuit of a large squeezing typically necessitates a strong nonlinearity, which, however, inevitably introduces additional decoherence and limits the accessibility of the large Hilbert space of an oscillator with a linear displacement drive. 

In the realm of quantum optics with superconducting microwave circuits~\cite{gu2017}, Kerr nonlinearity can be achieved using Josephson junctions, which are essential elements for developing parametric amplifiers~\cite{bergeal2010,macklin2015}, generating itinerant squeezed microwave fields~\cite{yurke1988,qiu2023}, and realizing superconducting qubits~\cite{blais2021}. By coupling to a strong nonlinear superconducting qubit, a harmonic oscillator in the single-photon Kerr regime has been demonstrated to realize the collapse and revival of coherent states~\cite{kirchmair2013} and to autonomously generate microwave Fock states~\cite{holland2015,li2024}. In addition, the use of weak Kerr nonlinearities has recently been proposed for manipulating photon-blockaded states~\cite{lingenfelter2021,ming2023}, with the operation speed enhanced by a large displacement. A similar idea was also demonstrated to implement fast conditional operations~\cite{eickbusch2022, diringer2024}. However, despite these theoretical and experimental advancements, demonstrating the fast generation of large squeezed states with a weak Kerr nonlinear oscillator remains a formidable experimental challenge.

In this work, we experimentally demonstrate the generation and amplification of squeezed states in a superconducting microwave cavity with weak Kerr nonlinearity. By subtly engineering an off-resonant microwave drive on the Kerr oscillator, we experimentally observe the cyclic dynamics of the squeezing evolution in a displaced frame without state collapse during the evolution. Additionally, we employ the Trotterization technique to realize quantum squeezing amplification by alternately displacing the Kerr oscillator, achieving a maximum squeezing degree of 14.6 dB, which, to our knowledge, is the largest squeezing value for microwave photonic states within the cavity. This technique offers an efficient approach for realizing displacement-enhanced squeezing operations to generate large squeezed states. Furthermore, we demonstrate the metrological advantages of these nonclassical states for sensing small displacements, promising potential applications in quantum-enhanced metrology.

\smallskip{}

\noindent \textbf{\large{}Results}{\large\par}

\begin{figure}
    \includegraphics{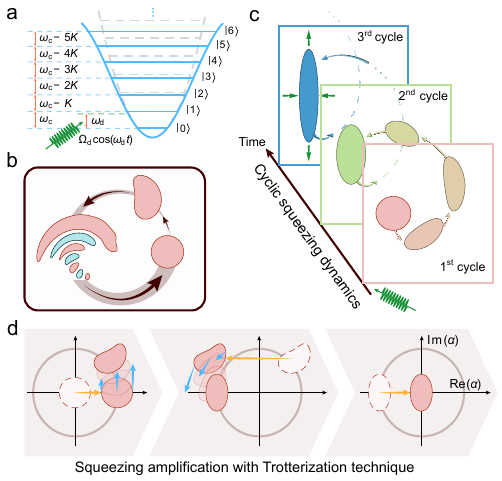}
    \caption{{Schematic illustration for achieving quantum squeezing amplification with a Kerr nonlinear oscillator.} 
    \textbf{a} Energy level diagram of a linear harmonic oscillator (gray dashed lines) and a Kerr nonlinear oscillator (solid blue lines) with an engineered detuned drive of frequency $\omega_\mathrm{d}$ and amplitude $\Omega_\mathrm{d}$ (green wavy line).
    \textbf{b} The quantum evolution of a coherent state represented with Wigner functions in phase space, exhibiting squeezing and collapse without any drives on the Kerr oscillator.
    \textbf{c} The quantum evolution of a coherent state to realize cyclic squeezing amplification with an engineered detuned drive on the Kerr oscillator. The coherent state undergoes a cyclic evolution in the phase space and becomes a squeezed state at the final moment of each cycle, where the squeezing degree gradually increases with the number of cycles.
    \textbf{d} The quantum evolution in phase space with Trotterization technique using displacement operations (yellow arrows) and Kerr-nonlinear rotations (blue arrows), showing the transition from a coherent state to a squeezed state.}
    \label{fig1}
\end{figure}

\noindent A linear harmonic oscillator has a quadratic potential well, which can be quantized into discrete and equally spaced energy levels, as depicted in Fig.~\ref{fig1}a. The oscillator would induce a self-Kerr nonlinearity when coupling to a nonlinear superconducting qubit in circuit quantum electrodynamics~\cite{blais2021}. This nonlinear Kerr oscillator has slightly unequally spaced energy levels with the Hamiltonian expressed as $H_K=-\frac{K}{2}{a^\dagger}^2a^2$ (assuming $\hbar=1$) in the rotating frame of the oscillator frequency $\omega_\mathrm{c}$. Here, $a$ ($a^\dagger$) is the annihilation (creation) operator, and $K$ is the strength of the Kerr nonlinearity of the oscillator. Under this Kerr Hamiltonian, a coherent state will first evolve into a slightly crooked squeezed state and then into a complete phase collapse state~\cite{kirchmair2013}, as indicated in Fig.~\ref{fig1}b. 

To avoid the state collapse and further improve the squeezing degree, we engineer an off-resonant microwave drive on the Kerr oscillator with a drive frequency of $\omega_\mathrm{d}$, resulting in a driven Kerr oscillator
\begin{eqnarray}
H_\mathrm{d} = \Delta_\mathrm{d} a^\dagger a -\frac{K}{2}{a^\dagger}^2a^2 + \Omega_\mathrm{d} \left(a+a^\dagger\right), \label{Hd}
\end{eqnarray}
where $\Delta_\mathrm{d} = \omega_\mathrm{c} - \omega_\mathrm{d}$ represents the frequency detuning between the oscillator and the drive, and $\Omega_\mathrm{d}$ is the strength of the drive (ignoring the drive phase for simplicity without any loss of generality). This Hamiltonian represents the well-known prototypical model of a quantum Duffing oscillator, exhibiting nonequilibrium phase transition dynamics~\cite{chen2023} and nonclassical Wigner negativity~\cite{Rosiek2024}.

The quantum dynamics of the driven Kerr Hamiltonian with engineered parameters can be employed to generate two-photon squeezing (Fig.~\ref{fig1}c), which can be further amplified through the Trotterization technique (Fig.~\ref{fig1}d). To intuitively understand the squeezing mechanism, we move into a displaced frame with a unitary operation of $D(\beta)=\exp{(\beta a^\dagger - \beta^* a)}$. Then, the aforementioned driven Kerr Hamiltonian can be transformed into
\begin{eqnarray}
H_{\beta}&=&\Delta'a^\dagger a-\frac{K}{2}{a^\dagger}^2a^2-\frac{K}{2}(\beta^2{a^\dagger}^2+\beta^{*2}{a}^2)\notag\\
&-&K\beta a^\dagger(a^\dagger a-r)+ \mathrm{H.C.} \label{Hbeta_main}
\end{eqnarray}
with $\Delta'=\Delta_\mathrm{d}-2K|\beta|^2$ and $r=\Delta_\mathrm{d}/K + \Omega_\mathrm{d}/K\beta - |\beta|^2$ representing the photon-blockade parameter. The first line of the above equation corresponds to the Hamiltonian of a Kerr parametric oscillator, which promises potential applications in quantum computation~\cite{Goto2016sci,puri2017,grimm2020,Frattini2024}, quantum metrology~\cite{Guo2024}, quantum phase transition~\cite{chavez2023, Zhang2024}, and quantum tunneling~\cite{Reynoso2023, Iyama2024}. 

\smallskip{}
\noindent \textbf{Cyclic squeezing dynamics}{\large\par}

\noindent Here we first discuss the driven squeezing dynamics. By properly engineering the detuned drive on the Kerr oscillator, we find that a coherent state no longer collapses during the evolution and undergoes a cyclic squeezing evolution, as shown in Fig.~\ref{fig1}c. Intuitively, the photon blockade term in the displaced frame arises from the interplay between the detuned drive and the Kerr nonlinearity. The detuned drive acts as a driving force pulling the coherent state along the negative imaginary axis in phase space, while the Kerr term rotates the state anticlockwise at a rate proportional to the average photon number. Carefully balancing these two effects prevents state collapse and enables periodic cycling trajectory dynamics (see Supplementary Figure~2). The mirror-symmetry trajectory around the real axis in phase space effectively cancels the photon blockade effect at the end of each cycle, achieving the displacement-enhanced squeezing operation with a squeezing rate $K\beta^2$, as indicated by the two-photon squeezing term in Eq.~(\ref{Hbeta_main}). The optimal parameters of the detuned drive are extracted from numerical simulations by evaluating the fidelity between the evolved squeezed state and an ideal squeezed state $|\xi\rangle = S(\xi)|0\rangle$, where $S(\xi) = \exp{(\frac{\xi^*}{2}a^2 - \frac{\xi}{2}{a^\dagger}^2 )}$ represents a squeezing operator. Here $\xi$ is a complex number of the squeezing parameter and $20\log_{10}{(e^{|\xi|})}$ defines the squeezing level or squeezing degree in dB. The details of the numerical simulations are presented in Supplementary Note~2.

\begin{figure*}
    \includegraphics{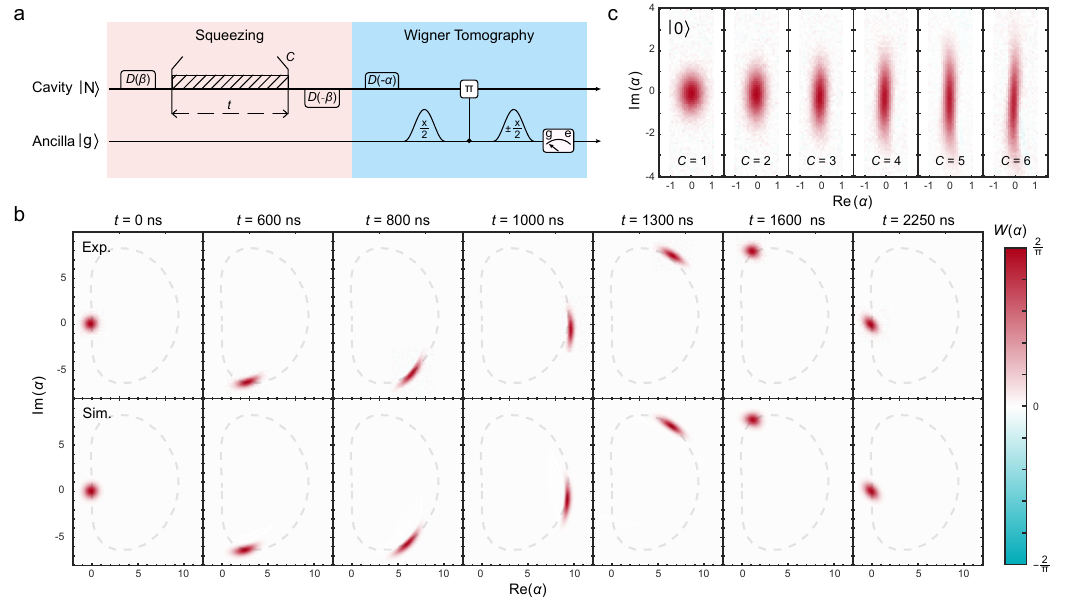}
    \caption{{Cyclic squeezing evolution with a detuned drive on the Kerr nonlinear oscillator for generating squeezed vacuum states.}
    \textbf{a} Experimental sequence for cyclic squeezing evolution and characterization with Wigner tomography.
    \textbf{b} Experimentally measured (top row) and numerically simulated (bottom row) Wigner function snapshots for the cyclic squeezing evolution of a vacuum state $|N=0\rangle$ in the first cycle (C=1), showcasing that the quantum state no longer collapses during the evolution and evolves into a squeezed state at the final moment of each cycle. 
    \textbf{c} Experimentally measured Wigner functions of the generated squeezed vacuum states at the final moment of the first 6 evolutionary cycles, obviously indicating that the squeezing is enhanced as the number of cycles increases. 
}
    \label{fig2}
\end{figure*}

In our experiment, we realize a Kerr nonlinear oscillator by dispersively coupling a high-quality superconducting microwave cavity (resonance frequency of $\omega_\mathrm{c}/2\pi=6.60$~GHz) to an ancillary superconducting qubit (transition frequency of $\omega_\mathrm{q}/2\pi=5.28$~GHz). The cavity mode has a weak Kerr nonlinearity of $K/2\pi=5.83$~kHz ($K/\omega_\mathrm{c}<10^{-6}$) and a single-photon lifetime of 395~$\mu$s (corresponding to a decay rate $\kappa_\mathrm{c}/2\pi = 0.40$~kHz), serving as the storage cavity for storing and manipulating the multiphoton squeezed states. The ancillary qubit mode, with an energy relaxation time of about 38~$\mu$s and a Ramsey (echo) decoherence time of 50 (58)~$\mu$s, is utilized for characterizing the generated squeezed states in the storage cavity. Detailed information regarding the device parameters and the experimental setup is provided in Supplementary Note~1. Note that, our displacement-enhanced squeezing method is also applicable even with an ultra-weak Kerr nonlinearity $K < \kappa_\mathrm{c}$ (see Supplementary Figure~6).

\begin{figure*}
    \includegraphics{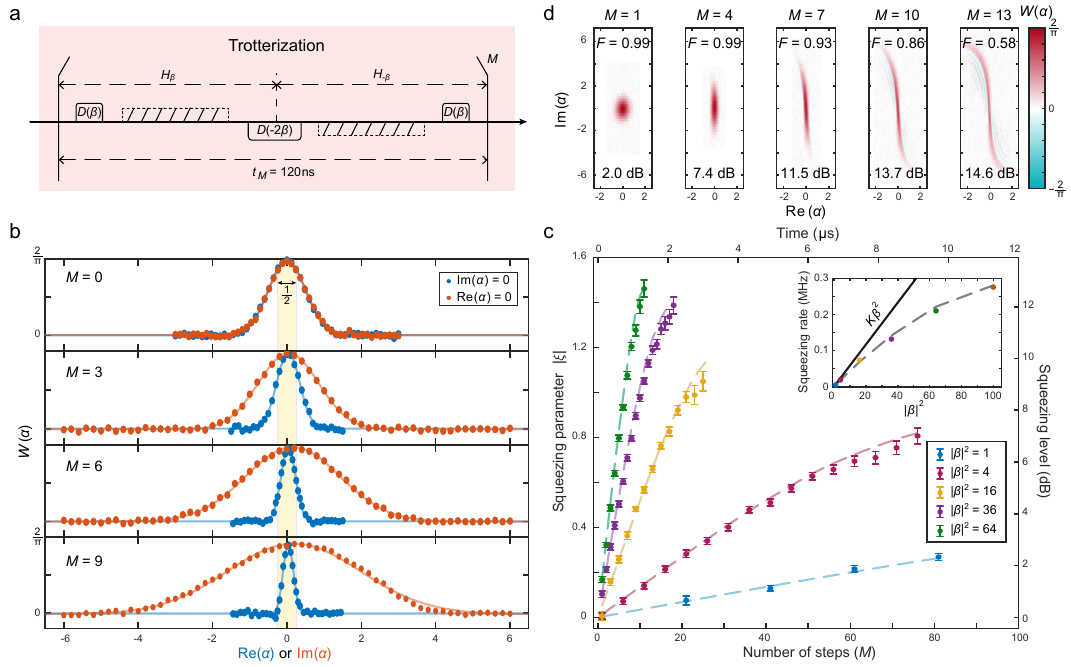}
    \caption{{Quantum squeezing amplification with the Trotterization technique.}
	\textbf{a} Experimental sequence of the Trotter steps for realizing the squeezing amplification. 
	\textbf{b} Measured 1D Wigner function cuts along the real axis (blue symbols) and the imaginary axis (red symbols) of the generated squeezed states for various Trotter steps. Solid lines are a global fit to these two traces at each Trotter step.
	\textbf{c} Experimentally extracted (symbols) and numerically simulated (dashed lines) squeezing parameters as a function of the number of Trotter steps (bottom axis) or the evolutionary time (top axis) for various displacement amplitudes $|\beta|^2$. Error bars are the estimated $95\%$ confidence intervals of the fittings. The inset exhibits the extracted squeezing rates as a function of the displacement amplitude $|\beta|^2$, which agree well with the simulation results (dashed line). Solid line represents the ideal squeezing rate of $K\beta^2$. 
	\textbf{d} Measured 2D Wigner functions of the generated squeezed states for various Trotter steps with $|\beta|^2=100$. The extracted state fidelities and squeezing levels are indicated accordingly.
}
    \label{fig3}
\end{figure*}

The experimental sequence is shown in Fig.~\ref{fig2}a, where the initial cavity vacuum state is first transformed into a displaced frame using a displacement operation $D(\beta)$, then evolves under the detuned driven Kerr Hamiltonian, and finally reverts to the original frame via a reverse displacement operation $D(-\beta)$. To achieve a larger squeezing degree, we choose an optimal displacement amplitude $\beta=2$, and drive frequency detuning $\Delta_\mathrm{d}/2\pi=56$~kHz and amplitude $\Omega_\mathrm{d}/2\pi=2.01$~MHz, as determined from numerical simulations (see Supplementary Figure~3). At the end of the sequence, we employ Wigner tomography to characterize the cavity states. During the first evolution cycle with a period of 2250~ns, we measure the Wigner functions of the cavity states at various moments, which are in good agreement with numerical simulations, as indicated in Fig.~\ref{fig2}b. The measurement results reveal that the coherent state rotates without collapsing in phase space under the detuned driven Kerr Hamiltonian, and is compressed into an elliptical structure in phase space at the end of each evolution cycle, indicating the successful generation of squeezed states. We displace the compressed coherent state to the origin in phase space, perform a virtual phase rotation operation to eliminate the rotation angle (see Supplementary Note~4), and measure the Wigner functions of the generated squeezed states at the final moment of different cycles. The experimental results displayed in Fig.~\ref{fig2}c, clearly show that as the number of cycles increases, the cavity state is gradually compressed in one direction and stretched in the orthogonal direction. This structural feature indicates that the squeezing level of the cavity state is progressively amplified with increasing cycles, demonstrating the effectiveness of the displacement-enhanced squeezing approach. 
In addition, the driven squeezing approach may offer potential applicability and flexibility in engineering complex squeezing evolution. For example, we employ this approach to generate squeezed multiphoton Fock states $S(\xi)|N\rangle$ with $N$ up to 6 (see Supplementary Figure~7). The generated squeezed Fock states, with a large squeezing level and photon number, could potentially achieve higher sensitivities than either squeezed vacuum states or multiphoton-number Fock states alone, offering significant advantages in quantum-enhanced precision metrology.

\smallskip{}
\noindent \textbf{Squeezing amplification with Trotterization technique}{\large\par}
\noindent To further enhance the squeezing levels without relying on brute-force optimizations, we propose utilizing the Trotterization technique~\cite{Lloyd1996} to eliminate the residual photon-blockade effect in Hamiltonian Eq.~(\ref{Hbeta_main}). This Trotterization strategy is based on the Kerr squeezing dynamics mentioned above, enabling a digital squeezing protocol that relies solely on simple, natively available gates and dynamics, as shown in Fig.~\ref{fig1}d. The key point of this approach lies in the addition of the driven Kerr Hamiltonian in two opposite displacement frames, $H_\beta$ and $H_{-\beta}$, which could completely eliminate the photon-blockade term, resulting in a Kerr parametric oscillator Hamiltonian:
\begin{eqnarray}
H_\mathrm{KPO} &=& \frac{H_\beta + H_{-\beta}}{2}\notag\\
               &=& \Delta' a^\dagger a-\frac{K}{2}{a^\dagger}^2a^2-\frac{K}{2}(\beta^2{a^\dagger}^2+\beta^{*2}{a}^2).
\label{HKPO_main}
\end{eqnarray}
Based on the Trotterization formula, the time evolution of the Kerr parametric oscillator Hamiltonian can be expressed as
\begin{eqnarray}
e^{-iH_\mathrm{KPO}\delta t} = e^{-iH_{-\beta} \delta t/2} e^{-iH_{\beta}\delta t/2} + \mathcal{O}[(\delta t)^2],
\end{eqnarray}
in a discretized evolutionary time step $\delta t$ with Trotter errors suppressed to an order of $\mathcal{O}[(\delta t)^2]$. This Trotterization method alternates the phases of the displacement frame, which effectively cancels the undesired nonlinear photon-blockade phase errors while reinforcing the two-photon squeezing term. By increasing the displacement amplitude $\beta$, the two-photon squeezing term would dominate in the Kerr parametric oscillator Hamiltonian of Eq.~(\ref{HKPO_main}) for achieving large squeezed states. Note that in this scenario, the rapid switching between opposite-signed displaced states prevents any single step of Kerr evolution from going far beyond the squeezing stages, rendering the detuned drive unnecessary during the Kerr evolution. We therefore set $\Omega_\mathrm{d}=0$ to restrict the average photon numbers of the intermediate states and engineer a virtual phase shift after each evolution cycle to mitigate the Kerr-induced phase on the squeezed state (see Supplementary Note~4).

\begin{figure} 
\includegraphics{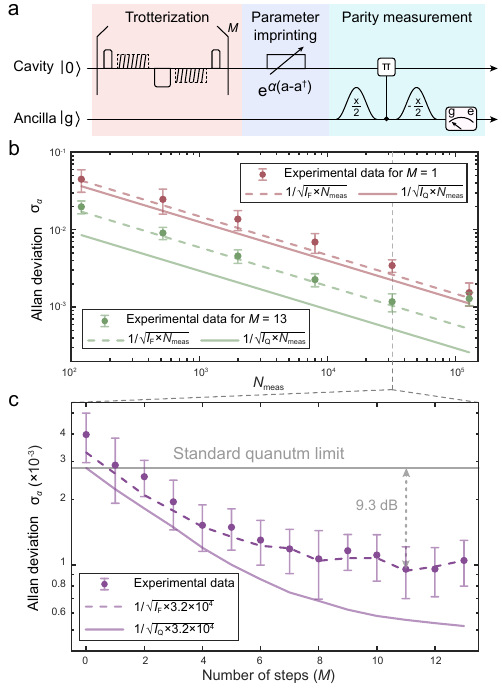}
\caption{{Quantum metrology for measuring small displacements using the generated squeezed states.} 
\textbf{a} Experimental sequence for the displacement measurement.
	\textbf{b} Allan deviation for estimating the displacement amplitude using squeezed states generated with different Trotter steps (red: $M=1$, green: $M=13$) versus the number of independent measurements $N_\mathrm{meas}$. The increase of the last point for $M=13$ may primarily be attributed to displacement drifts in the experimental apparatus.
	\textbf{c} Allan deviation for estimating the displacement amplitude as a function of Trotter steps for generating the squeezed states with a fixed $N_\mathrm{meas}=3.2\times10^4$. Solid (dashed) line represents the theoretical Cram\'er-Rao bound using quantum (classical) FI of the squeezed states. Error bars are standard deviations from repeated experiments. 
}
\label{fig4}
\end{figure}

The experimental sequence to demonstrate the quantum squeezing amplification with the Trotterization method is shown in Fig.~\ref{fig3}a. In this sequence, we alternately transform the Kerr oscillator into different displacement frames with opposite phases. This is similar to an echoed displaced Kerr evolution that would average away the undesired photon-blockade term and suppress cavity dephasing errors (see Supplementary Note~7). To evaluate the squeezing performance of the generated squeezed states, we measure one-dimensional (1D) Wigner function cuts along the real axis ($\mathrm{Im}(\alpha)=0$) and the imaginary axis ($\mathrm{Re}(\alpha)=0$), respectively, with the experimental results shown in Fig.~\ref{fig3}b. As the Trotter steps increase, the linewidth of one trace is compressed below $1/2$, while the other one is expanded beyond $1/2$, indicating an effective squeezing of the quantum state. From a global fit to both traces, we extract the squeezing parameters $|\xi|$ of the generated squeezed states as a function of the number of Trotter steps for various displacement amplitudes $|\beta|$, with the experimental results shown in Fig.~\ref{fig3}c. 

The experimental results reveal that as the number of Trotter steps increases, the squeezing level is amplified towards a large squeezed state. The eventual saturation in squeezing levels is attributed to Kerr nonlinearity, verified by numerical simulations. By performing a linear fit to the experimental data within the linear region, we extract the squeezing rate as a function of $|\beta|^2$, as shown in the inset of Fig.~\ref{fig3}c. As the displacement amplitude $|\beta|$ increases, the squeezing rate is enhanced accordingly, but deviates from the ideal relationship $K\beta^2$. This deviation is primarily due to the neglect of Kerr evolution during the displacement operation and the influence of higher-order Kerr nonlinearities. When these aspects are taken into account, the numerical simulation results align well with the experimental data.

Furthermore, we have measured the 2D Wigner functions of the generated states at various Trotter steps, as shown in Fig.~\ref{fig3}d. The 2D Wigner fitting results demonstrate a maximum squeezing degree of 14.6 dB, which is larger than previous demonstrations of intracavity squeezing~\cite{wang2017,dassonneville2021,eickbusch2022,eriksson2024}. Despite the gradual twisting and stretching of the squeezed states in phase space due to Kerr nonlinearity, these states continue to exhibit large Fisher information (FI), which quantifies the maximum information that can be extracted about a parameter from the quantum state.

\smallskip{}

\noindent \textbf{Quantum metrology for sensing small displacement}{\large\par}
\noindent To assess the practical utility of these generated squeezed states, we further conduct a quantum metrology experiment to measure small displacements along the position quadrature using these states. In this experiment, we employ a parity measurement to map the information about the displacement onto the qubit state, with the experimental sequence illustrated in Fig.~\ref{fig4}a. From the measured qubit ground state population $P_{g}$, the classical FI is calculated with the formula $I_\mathrm{F}(\alpha)=\frac{1}{P_{g}(1-P_{g})}(\frac{\mathrm{d}P_{g}}{\mathrm{d}\alpha})^2$ and is bounded by quantum FI $I_\mathrm{Q}=4e^{2|\xi|}$ for ideal squeezed state $|\xi\rangle$. These FIs are explicitly linked to the variance of the estimated displacement amplitude via the Cram\'er-Rao inequality~\cite{PARIS2009}
\begin{eqnarray}
\Delta\alpha\ge\frac{1}{\sqrt{N_\mathrm{meas}I_\mathrm{F}}}\ge\frac{1}{\sqrt{N_\mathrm{meas}I_\mathrm{Q}}},
\end{eqnarray}
where $N_\mathrm{meas}$ represents the number of independent measurements.

Using the parity measurement results from $N_\mathrm{meas}$ experiments, we estimate a sample of displacement amplitude $\alpha$ by leveraging the Bayesian inference (see Supplementary Note~5). The variance of the estimated displacement amplitude is calculated by the Allan deviation formula $\sigma_\alpha=\sqrt{\frac{1}{2(l-1)}\sum_{k=1}^{l-1}(\alpha_{k+1}-\alpha_k)^2}$ with $l=10$ samples. In Fig.~\ref{fig4}b, we plot the Allan deviation for displacement estimation versus $N_\mathrm{meas}$ using the squeezed states generated with different Trotter steps. Additionally, Fig.~\ref{fig4}c depicts the measured Allan deviation as a function of the number of Trotter steps with a fixed $N_\mathrm{meas}=3.2\times10^4$. The results indicate that as the number of Trotter steps increases, the Allan deviation for displacement measurement significantly decreases, highlighting the large metrological advantage of these nonclassical states in surpassing the standard quantum limit. 

\smallskip{}

\noindent \textbf{\large{}Discussion}{\large\par}
\noindent In conclusion, we have experimentally demonstrated displacement-enhanced quantum squeezing amplification with a weak Kerr nonlinear oscillator in a superconducting microwave cavity, leveraging the Trotterization technique. This method distinguishes itself from previous displacement-enhanced operations~\cite{eickbusch2022, diringer2024} and other squeezing approaches~\cite{wang2017,dassonneville2021,eickbusch2022,qiu2023,eriksson2024} by eliminating decoherence errors induced by the ancillary qubit and suppressing cavity dephasing errors through echoed displacements during the evolution. Our deterministic squeezing approach ultimately generates nonclassical microwave photonic states with a maximum squeezing degree of 14.6 dB over quasiclassical coherent states. The squeezing degree is primarily limited by the displacement drive amplitude due to electronic hardware constraints and the breakdown of the dispersive approximation (see Supplementary Note~6). These large squeezed states are demonstrated to exhibit large metrological advantages, promising potential applications in quantum-enhanced metrology~\cite{Guo2024}. Furthermore, our approach provides a hardware-efficient way to engineer strong two-photon squeezing operations, which is beneficial for quantum information processing with Kerr cat qubits~\cite{Goto2016sci,puri2017,grimm2020}, bosonic quantum chemistry simulations~\cite{dutta2024}, and topological phase transitions~\cite{chavez2023, Zhang2024}, as well as for reducing decoherence of quantum systems~\cite{murch2013,jeannic2018,pan2023}. A recent notable advancement in quantum squeezing with acoustic wave resonators~\cite{marti2024} suggests the potential for easily adapting our approach to achieve even larger squeezing with microwave phonons, photons, or magnons in qubit-oscillator systems~\cite{wollman2015,yutaka2015,vonlupke2022,liu2024}.

\vbox{}
\smallskip{}

\noindent \textbf{\large{}Methods}{\large\par}

\noindent \textbf{Characterization of squeezing parameters}

\noindent The generated squeezed states are characterized by comparing measured Wigner functions to those of ideal squeezed states $|\xi\rangle$. The theoretical 2D Wigner function for an ideal squeezed state is given by $W(\alpha)=\frac{2}{\pi}\exp{(-2|\nu|^2)}$, where $\nu=\cosh(|\xi|)\alpha^*+\sinh(|\xi|)\alpha$ and $\alpha$ is the complex phase-space variable. The measured 2D Wigner functions in Figs.~\ref{fig3}d and \ref{fig2}b-c are fitted directly with this expression to extract the squeezing parameter. However, measuring the full 2D Wigner function is time-consuming. Therefore, we employ an efficient approach by measuring only 1D Wigner cuts along the squeezing quadrature ($\mathrm{Im}(\alpha)=0$) and the anti-squeezing quadrature ($\mathrm{Re}(\alpha)=0$), then performing a global fit to these two traces to efficiently extract the squeezing parameter $|\xi|$ (see Fig.~\ref{fig3}b). See the Supplementary Note 3 for detailed descriptions.

\vbox{}

\noindent \textbf{Extraction of Fisher information} 

\noindent In quantum metrology experiments, we employ the FI to quantify the precision achievable in estimating the displacement parameter $\alpha$ using nonclassical squeezed states. The classical FI is defined as $I_\mathrm{F}(\alpha) = \sum_\mu {\frac{1}{P_\mu} \left( \frac{\partial  P_\mu}{\partial \alpha}  \right)^2 }$~\cite{PARIS2009}, where $P_\mu$ denotes the probability distribution from a set of measurement projections ${M_\mu}$. In our experiment, the measurement performed on the ancillary qubit yields only two outcomes, $\mu=\{g,e\}$, corresponding to the qubit in ground and excited states. Consequently, the classical FI simplifies to $I_\mathrm{F}(\lambda) = \frac{1}{P_g(1-P_g)} \left( \frac{d P_g}{d\lambda} \right)^2$. For an ideal squeezed state $|\xi\rangle$, the quantum FI is calculated by $I_Q = 4(\langle \xi | h_\alpha^2  |\xi\rangle - \langle \xi | h_\alpha | \xi \rangle ^2) = 4e^{2|\xi|}$, with $h_\alpha = i(a - a^\dag)$ representing the interrogation Hamiltonian that encodes the displacement parameter via the unitary transformation $U_\alpha = \exp{(-i\alpha h_\alpha)}$ (assuming $\alpha$ is real). These FI quantities are directly related to the variance of the estimated displacement amplitude through the Cram\'er-Rao inequality~\cite{PARIS2009}: $\Delta\alpha\ge\frac{1}{\sqrt{N_\mathrm{meas}I_\mathrm{F}}}\ge\frac{1}{\sqrt{N_\mathrm{meas}I_\mathrm{Q}}}$. 

\vbox{}

\noindent \textbf{Bayesian inference} 

\noindent To estimate the displacement parameter $\alpha$ from the measured qubit population, we employ the Bayesian inference approach by maximizing the posterior probability $P(\alpha|\mathcal{M})\propto P(\alpha)\times P(\mathcal{M}|\alpha)$, where $P(\mathcal{M}|\alpha)$ is the likelihood of observing the ancilla in the ground state ($\mathcal{M}=g$) or excited state ($\mathcal{M}=e$) under $\alpha$. Assuming a uniform prior probability $P(\alpha)$ over $\alpha\in[0,0.5]$, Bayesian estimation reduces to maximum likelihood estimation. The displacement amplitude is estimated by maximizing the log-likelihood function $L(\mathcal{M}|\alpha) = \ln[P(\mathcal{M}|\alpha)]=P_g^{\mathrm{exp}}\ln[P_g(\alpha)] + (1-P_g^{\mathrm{exp}})\ln[1-P_g(\alpha)]$ via $\alpha_{\mathrm{est}}=\arg\max_\alpha L(\mathcal{M}|\alpha)$, incorporating the measured qubit ground state population $P_g^{\mathrm{exp}}$ and fitted qubit ground state probability distribution $P_g(\alpha)$ (see Supplementary Note 5). 

\vbox{}

\smallskip{}

\noindent \textbf{\large{}Data availability}{\large\par}

\noindent Source data are provided with this paper, and more detailed data are available from the corresponding authors upon request.  
\smallskip{}

\vbox{}
\noindent \textbf{\large{}Code availability}{\large\par}

\noindent The code used for this study is available from the corresponding authors upon request.

\vbox{}

%

\smallskip{}

\vbox{}
\noindent \textbf{\large{}Acknowledgment}{\large\par}

\noindent We thank Zhen-Biao Yang and Xiu Gu for valuable suggestions related to this work. This work was supported by the Quantum Science and Technology-National Science and Technology Major Project (Grants No.~2024ZD0302300, No.~2021ZD0301703), the Guangdong Basic and Applied Basic Research Foundation (Grant No.~2024B1515020013), the National Natural Science Foundation of China (Grants No.~12422416, No.~12274198, No.~12575025), the Shenzhen Science and Technology Program (Grant No.~RCYX20210706092103021), the Guangdong Provincial Key Laboratory (Grant No.~2019B121203002), the Shenzhen-Hong Kong cooperation zone for technology and innovation (Contract No.~HZQB-KCZYB-2020050).

\smallskip{}

\noindent \textbf{\large{}Author contributions}{\large\par}

\noindent Y.X. conceived the idea and designed the experiment. Y.C. and X.D. performed the experiments, data analysis, and numerical simulations under the supervision of Y.X. L.Z. fabricated the superconducting qubit chip assisted by S.L. Z.N., J.M., P.Z., and L.H. contributed to the experimental setup and Y.C., X.D., P.H., and Y.X. discussed the results. Y.C. and Y.X. wrote the manuscript with input from all authors. Y.X. and D.Y. supervised the project. 

\smallskip{}

\noindent \textbf{\large{}Competing interests}{\large\par}

\noindent The authors declare no competing interests.

\smallskip{}

\clearpage{}

\end{document}


\title{Supplementary Information for ``Quantum squeezing amplification with a weak Kerr nonlinear oscillator"}

\author{Yanyan Cai}
\thanks{These authors contributed equally to this work.}
\affiliation{Southern University of Science and Technology, Shenzhen 518055, China}
\affiliation{International Quantum Academy, Shenzhen 518048, China}

\author{Xiaowei Deng}
\thanks{These authors contributed equally to this work.}
\affiliation{International Quantum Academy, Shenzhen 518048, China}

\author{Libo Zhang}
\thanks{These authors contributed equally to this work.}
\affiliation{Southern University of Science and Technology, Shenzhen 518055, China}
\affiliation{International Quantum Academy, Shenzhen 518048, China}

\author{Zhongchu Ni}
\affiliation{International Quantum Academy, Shenzhen 518048, China}

\author{Jiasheng Mai}
\affiliation{Southern University of Science and Technology, Shenzhen 518055, China}
\affiliation{International Quantum Academy, Shenzhen 518048, China}

\author{Peihao Huang}
\affiliation{International Quantum Academy, Shenzhen 518048, China}

\author{Pan Zheng}
\affiliation{International Quantum Academy, Shenzhen 518048, China}

\author{Ling Hu}
\affiliation{International Quantum Academy, Shenzhen 518048, China}
\affiliation{Shenzhen Branch, Hefei National Laboratory, Shenzhen 518048, China}
\author{Song Liu}
\affiliation{International Quantum Academy, Shenzhen 518048, China}
\affiliation{Shenzhen Branch, Hefei National Laboratory, Shenzhen 518048, China}

\author{Yuan Xu}
\email{xuyuan@iqasz.cn}
\affiliation{International Quantum Academy, Shenzhen 518048, China}
\affiliation{Shenzhen Branch, Hefei National Laboratory, Shenzhen 518048, China}

\author{Dapeng Yu}
\affiliation{International Quantum Academy, Shenzhen 518048, China}
\affiliation{Shenzhen Branch, Hefei National Laboratory, Shenzhen 518048, China}

\maketitle
\vskip 0.5cm


\begin{nonumsection}
    \section{Supplementary Note \thesection~-~Experimental device and setup}
\end{nonumsection}
We experimentally demonstrate quantum squeezing amplification in a three-dimensional (3D) circuit quantum electrodynamics (QED) architecture~\cite{blais2021}. The experimental device is similar to that in Refs.~\cite{ni2023,deng2024} and comprises three key components: a 3D coaxial stub cavity~\cite{reagor2016}, a fixed-frequency superconducting transmon qubit~\cite{koch2007}, and a Purcell-filtered stripline readout resonator~\cite{axline2016}, as illustrated in Supplementary Figure~\ref{setup}. 

The 3D circuit QED device is directly machined from a high-purity (5N5) aluminum block and undergoes chemical etching to extend the cavity's coherence lifetime~\cite{reagor2013}. The coaxial stub cavity~($C$) is designed as a 3D $\lambda/4$ transmission line resonator, whose fundamental mode is utilized as the storage cavity for storing the generated squeezed states. The superconducting transmon qubit~($Q$), fabricated on a sapphire chip with two antenna pads, serves as an auxiliary qubit. One pad couples to the storage cavity, while the other couples to the stripline readout resonator~($R$). The coupling to the nonlinear qubit induces a weak Kerr nonlinearity to the storage cavity, effectively realizing a Kerr nonlinear oscillator. The Purcell-filtered readout resonator consists of two planar $\lambda/2$ striplines: one is strongly coupled to the transmon qubit for fast dispersive readout of qubit states, and the other is coupled to the external environment, acting as a Purcell filter to preserve the coherence lifetimes of both the qubit and the storage cavity.

The experimental device is placed inside a magnetic shield in a cryogen-free dilution refrigerator at a temperature below $10$~mK. The control wiring of the system is similar to that in Ref.~\cite{ni2022}. All microwave control pulses for the auxiliary qubit, storage cavity, and readout resonator are generated using single-sideband in-phase and quadrature (IQ) modulations with an arbitrary waveform generator~(Tektronix AWG5208). These signals are transmitted to the experimental device through coaxial cables with microwave isolators, filters, and attenuators to minimize reflection waves and radiation noise. The transmitted readout signal is amplified by a high electron mobility transistor (HEMT) amplifier at 4K stage, followed by a standard commercial radio frequency (RF) amplifier at room temperature. Subsequently, the signal is downconverted using a microwave mixer with the same local oscillator (LO) frequency as that used to generate the readout pulse and is recorded and digitized by a data acquisition card~(AlazarTech ATS9870) along with corresponding reference signals. Detail information on the wiring and experimental setup is provided in Supplementary Figure~\ref{setup}.

\begin{figure}[htbp]
  \centering
  \includegraphics{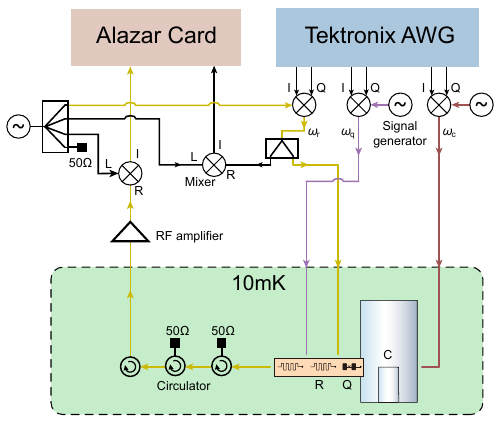}
  \caption{Control wiring and experimental setup.}
  \label{setup}
\end{figure}

\begin{nonumsection}
    \section{Supplementary Note \thesection~-~Quantum dynamics of the detuned driven Kerr oscillator}
\end{nonumsection}

\subsection{System Hamiltonian and parameters}
In this 3D circuit QED system, the superconducting qubit is dispersively coupled to both the storage cavity and the readout resonator, with the dispersive Hamiltonian expressed as:
\begin{eqnarray}
H/\hbar &=& \omega_\mathrm{c} a^\dagger a -\frac{K}{2}{a^\dagger}^2a^2\notag\\
&+&\omega_\mathrm{q} q^\dagger q -\frac{\eta_\mathrm{q}}{2}{q^\dagger}^2q^2-\chi_\mathrm{qc}a^\dagger aq^\dagger q\notag\\
&+&\omega_\mathrm{r} a_\mathrm{r}^\dagger a_\mathrm{r}-\chi_\mathrm{qr}a_\mathrm{r}^\dagger a_\mathrm{r}q^\dagger q,
\label{H_total}
\end{eqnarray}
where $a$,$q$ and $a_\mathrm{r}$ are the annihilation operators for the storage cavity, superconducting qubit, and readout resonator, respectively; $\omega_\mathrm{c}$, $\omega_\mathrm{q}$, and $\omega_\mathrm{r}$ are their respective resonance frequencies; $K$ is the Kerr nonlinearity of the storage cavity; $\eta_\mathrm{q}$ is the anharmonicity of the qubit; and $\chi_\mathrm{qc}$ and $\chi_\mathrm{qr}$ are the cross-Kerr interactions between the qubit and the storage cavity and readout resonator, respectively. The measured values of these parameters are summarized and listed in Supplementary Table~\ref{TableS1}. 

\begin{table}
\caption{Device parameters of the system.}
\begin{tabular}{p{7cm}<{\raggedright} p{1.2cm}<{\centering}}
  \hline
  \hline
  Parameters & Value \\\hline
  &\\[-0.9em]
  Storage cavity frequency $\omega_\mathrm{c}/2\pi$(GHz) & 6.60 \\
  &\\[-0.9em]
  Storage cavity Kerr nonlinearity $K/2\pi$(kHz) & 5.83\\
  &\\[-0.9em]
  Storage cavity dispersive shift $\chi_\mathrm{qc}/2\pi$(MHz) & 1.94\\
  &\\[-0.9em]
  Storage cavity relaxation $T_\mathrm{1,c}$ ($\mu$s) & 395\\
  &\\[-0.9em]
  Storage cavity Ramsey coherence $T_\mathrm{2,c}$ ($\mu$s) & 595\\
  &\\[-0.9em]
  Storage cavity thermal population & 0.1\%\\
  &\\[-0.9em]
  Qubit frequency $\omega_\mathrm{q}/2\pi$(GHz) & 5.28 \\
  &\\[-0.9em]
  Qubit anharmonicity $\eta_\mathrm{q}/2\pi$(MHz) & 194.6 \\
  &\\[-0.9em]
  Qubit relaxation $T_\mathrm{1,q}$ ($\mu$s) & 38\\
  &\\[-0.9em]
  Qubit Ramsey coherence $T_\mathrm{2,q}$ ($\mu$s) & 50\\
  &\\[-0.9em]
  Qubit echo coherence $T_\mathrm{2E,q}$ ($\mu$s) & 58\\
  &\\[-0.9em]
  Qubit thermal population & 1.15\%\\
  &\\[-0.9em]
  Readout frequency $\omega_\mathrm{r}/2\pi$(GHz) & 8.67 \\
  &\\[-0.9em]
  Readout decay rate $\kappa_\mathrm{r}/2\pi$(MHz) & 3.8\\
  &\\[-0.9em]
  Readout dispersive shift $\chi_\mathrm{qr}/2\pi$(MHz) & 2.0\\
  &\\[-0.9em]
  \hline
  \hline
\end{tabular}
\label{TableS1}
\end{table}

\subsection{Quantum dynamics in phase space}
Focusing solely on the storage cavity mode, the above Hamiltonian simplifies to a Kerr nonlinear Hamiltonian $H_K=\omega_\mathrm{c}a^\dagger a-\frac{K}{2}{a^\dagger}^2a^2$ (assuming $\hbar=1$). Under this Hamiltonian, a coherent state evolves into a slightly squeezed state and then collapses in phase space~\cite{kirchmair2013}. 
By applying an engineered off-resonant microwave drive $\Omega_\mathrm{d}(e^{i\omega_\mathrm{d}t}a+e^{-i\omega_\mathrm{d}t}a^\dagger)$, the coherent state undergoes a cyclic squeezing dynamics evolution in phase space without collapse. Here, $\Omega_\mathrm{d}$ and $\omega_\mathrm{d}$ denote the drive strength and frequency, respectively. In the rotating frame of the oscillator frequency $\omega_\mathrm{c}$, the driven Kerr Hamiltonian is expressed as:
\begin{eqnarray}
H_\mathrm{d} = \Delta_\mathrm{d} a^\dagger a -\frac{K}{2}{a^\dagger}^2a^2 + \Omega_\mathrm{d} \left(a+a^\dagger\right),
\label{Hd}
\end{eqnarray}   
where $\Delta_\mathrm{d} = \omega_\mathrm{c} - \omega_\mathrm{d}$ is the frequency detuning between the oscillator and the drive.

\begin{figure}[b]
  \centering
  \includegraphics{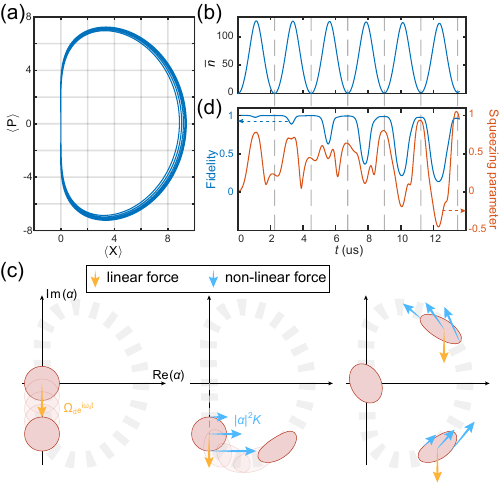}
  \caption{Cyclic evolution of the driven Kerr nonlinear oscillator. 
  \textbf{(a)} Quantum state trajectories for the first six evolutionary cycles in phase space.
  \textbf{(b)} Time evolution of the average photon number of the quantum state in the cavity.
  \textbf{(c)} Schematic illustration for the quantum dynamics of the detuned driven Kerr oscillator. Grey dashed lines are the evolutionary trajectories of the simulation in (a).
  \textbf{(d)} Time evolution of the squeezing parameters and fidelity between the evolved state and an ideal squeezed vacuum state. Dashed vertical lines correspond to the final moment of each evolution cycle.}
  
  \label{cycleT_sim}
\end{figure} 

To investigate the cyclic quantum dynamics behavior of the detuned driven Kerr Hamiltonian, we numerically simulate the evolution of a coherent state, which is initialized using a displacement operator $D(\beta)=e^{\beta a^{\dagger}-\beta^{*}a}$, under the Hamiltonian $H_\mathrm{d}$ by solving the master equation. The simulated cavity state is finally displaced back to the origin in phase space using a reverse displacement operator $D(-\beta)$. In addition to the Hamiltonian in Supplementary Equation~(\ref{Hd}), higher-order Kerr nonlinear effects with strengths on the order of a few hertz are also considered in the simulations to ensure accuracy, although they are not explicitly shown here. From the simulation, we extract the expectation values of quadrature operators $X=(a+a^{\dagger})/2$ and $P=-i(a-a^{\dagger})/2$ during the evolution and reconstruct the quantum state trajectory of the cavity state in phase space. As shown in Supplementary Figure~\ref{cycleT_sim}(a), the trajectory exhibits cyclic evolution, and the average photon number $\langle a^\dagger a\rangle$, depicted in Supplementary Figure~\ref{cycleT_sim}(b), also displays periodic oscillations. This cyclic dynamical evolution can be illustrated in Supplementary Figure~\ref{cycleT_sim}(c), where the microwave drive acts as a global linear force on the quantum state in phase space, pulling it along the negative imaginary axis. Meanwhile, the Kerr nonlinearity in the system introduces an amplitude-dependent non-linear force, which rotates the state around the origin of phase space anticlockwise with an angular velocity $|\alpha|^2K$. This rotational frequency increases quadratically with the distance from the origin. As a result of these combined forces, the quantum state undergoes periodic evolution in phase space. The state trajectory maintains mirror symmetry about the real axis in phase space, which helps to cancel out undesired distortions. This process enables the transformation from a slightly curved, banana-shaped state during intermediate stages to an ideal squeezed state at the end of the evolution cycle. To verify this point, we numerically investigate both the state fidelity and squeezing parameter as a function of evolution time in Supplementary Figure~\ref{cycleT_sim}(d). The results indicate that the fidelity reaches a maximum at the endpoint of the evolution cycle, and the squeezing parameter increases progressively with the number of cycles.

\begin{figure}[tb]
  \centering
  \includegraphics{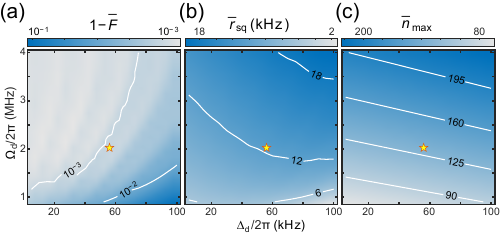}
  \caption{Numerical simulated average infidelity of the generated squeezed states \textbf{(a)}, average squeezing rate \textbf{(b)}, and maximum average photon number \textbf{(c)} during the evolution of the cavity state as a function of the frequency detuning $\Delta_\mathrm{d}=\omega_\mathrm{c}-\omega_\mathrm{d}$ and the drive amplitude $\Omega_\mathrm{d}$ for the first 6 cycles. The star marks the optimal choice of drive parameters for generating the squeezed Fock states in our experiment.}
  \label{sim_scan}
\end{figure}

To determine the optimal driving parameters for achieving a larger squeezing level, we perform numerical simulations to investigate the performance of the evolved squeezed states after the first six evolution cycles under the Hamiltonian $H_\mathrm{d}$, while scanning the frequency detuning $\Delta_\mathrm{d}$ and the strength $\Omega_\mathrm{d}$. In the simulation, the cavity is initialized in a coherent state $|\beta=2\rangle$. Here, the choice of $\beta=2$ represents a trade-off that balances the squeezing performance and higher-order Kerr nonlinear effects. A smaller displacement amplitude $\beta$ results in an insufficient squeezing rate, whereas a larger $\beta$ amplifies higher-order Kerr nonlinearities, which can deform the squeezed states. We evaluate the average infidelity $1-\bar{F}$ of the evolved squeezed states after the first six evolution cycles compared to ideal squeezed states as a function of $\Delta_\mathrm{d}$ and $\Omega_\mathrm{d}$, with the results shown in Supplementary Figure~\ref{sim_scan}(a). The average squeezing rate $\bar{r}_\mathrm{sq}$ is also calculated using the squeezing parameters and cyclic periods of the evolution, as shown in Supplementary Figure~\ref{sim_scan}(b). Additionally, we extract the maximum average photon number $\bar{n}_\mathrm{max}$ of the cavity states during the cyclic squeezing evolution, as shown in Supplementary Figure~\ref{sim_scan}(c). The simulation results indicate that within the parameter space, higher squeezing rates and fidelities are associated with larger photon numbers during the evolution, which, however, can break the dispersive approximation between the ancillary qubit and the cavity \cite{blais2021}, similar to the case discussed in Supplementary Note~\ref{breakdown}. The optimal driving parameters used in the experiment are chosen by balancing the above three physical quantities $\bar{n}_\mathrm{max}$, $r_{\mathrm{sq}}$ and $\bar{F}$ in the simulation. As a result, we choose the drive frequency detuning of $\Delta_\mathrm{d}/2\pi=56$kHz and drive strength of $\Omega_\mathrm{d}/2\pi=2.01$MHz (marked by a star in Supplementary Figure~\ref{sim_scan}), which yield a squeezing fidelity $\bar{F}=99.89\%$ and an average squeezing rate $r_\mathrm{sq}/2\pi=12.22$~kHz after the first six evolution cycles in numerical simulation.

\subsection{Analysis of the cycling evolutionary period}

\subsubsection{Semiclassical approximation}

The cyclic dynamical evolution demonstrated in Supplementary Figure~\ref{cycleT_sim} can be analyzed through the periodicity revealed by the Heisenberg equations of motion for $X$ and $P$, that is, $\mathrm{d}X/\mathrm{d}t=-i[X,H_d]$, $\mathrm{d}P/\mathrm{d}t=-i[P,H_d]$. Based on Ehrenfest's theorem and employing a semiclassical approximation, the differential equations can be reformulated as:
\begin{eqnarray}
\frac{\mathrm{d}x}{\mathrm{d}t}&=&\Delta_dp-Kp(p^2+x^2),\notag\\
\frac{\mathrm{d}p}{\mathrm{d}t}&=&-\Delta_dx+Kx(p^2+x^2)-\Omega_d,
\label{OED_semi}
\end{eqnarray}
where $x=\langle X\rangle$ and $p=\langle P\rangle$.

Through dimensionless treatment with  $\epsilon=\Delta_\mathrm{d}/\Omega_\mathrm{d}$, $k=K/\Omega_\mathrm{d}$ and $\tau=t\Omega_\mathrm{d}$, Supplementary Equations~(\ref{OED_semi}) can be reformulated as  
\begin{eqnarray}
\frac{\mathrm{d}x}{\mathrm{d}\tau}&=&\epsilon p-kp(p^2+x^2),\notag\\
\frac{\mathrm{d}p}{\mathrm{d}\tau}&=&-\epsilon x+kx(p^2+x^2)-1.
\label{OED_semi2}
\end{eqnarray}
Given that $\Delta_\mathrm{d} \ll \Omega_\mathrm{d}$ such that $\epsilon\ll 1$, we assume that the solution of Supplementary Equations~(\ref{OED_semi2}) can be expanded in a perturbative series as 
\begin{eqnarray}
x(\tau)&=&x_0(\tau)+\epsilon x_1(\tau)+\mathcal{O}(\epsilon^2),\notag\\
p(\tau)&=&p_0(\tau)+\epsilon p_1(\tau)+\mathcal{O}(\epsilon^2),
\label{series_pertur}
\end{eqnarray}
where $x_0$, $x_1$, $p_0$ and $p_1$ are all independent of $\epsilon$. Substituting Supplementary Equations~(\ref{series_pertur}) into Supplementary Equations~(\ref{OED_semi2}) and retaining the zeroth-order and first-order terms in $\epsilon$, we obtain
\begin{eqnarray}
\frac{\mathrm{d}x_0}{\mathrm{d}\tau}&=&-kp_0(p_0^2+x_0^2),\notag\\
\frac{\mathrm{d}p_0}{\mathrm{d}\tau}&=&kx_0(p_0^2+x_0^2)-1,
\label{ODE_pertur0}
\end{eqnarray}
and
\begin{eqnarray}
\frac{\mathrm{d}x_1}{\mathrm{d}\tau}&=&p_0-2kp_0x_0x_1-kp_1(3p_0^2+x_0^2),\notag\\
\frac{\mathrm{d}p_1}{\mathrm{d}\tau}&=&-x_0+2kx_0p_0p_1+kx_1(p_0^2+3x_0^2).
\label{ODE_pertur1}
\end{eqnarray}
Solving Supplementary Equations~(\ref{ODE_pertur0}), we obtain a constant solution $x_0=k^{-1/3}$, $p_0=0$. Substituting these into Supplementary Equations~(\ref{ODE_pertur1}) yields a general solution as:
\begin{eqnarray}
x_1(\tau)&=&\mathcal{C}_1\cos(\omega_\tau \tau)+\frac{1}{3k^{2/3}},\notag\\
p_1(\tau)&=&\mathcal{C}_2\sin(\omega_\tau\tau),
\label{solution}
\end{eqnarray}
where $\omega_\tau=\sqrt{3}k^{1/3}$, and $\mathcal{C}_1$ and $\mathcal{C}_2$ are undetermined coefficients related to the initial conditions. Consequently, the evolutionary period is given by 
\begin{eqnarray}
T_c=2\pi/(\omega_\tau\Omega_d)=2\pi/(\sqrt{3}K^{1/3}\Omega_d^{2/3}).
\end{eqnarray}

\begin{figure}[b]
  \centering  \includegraphics{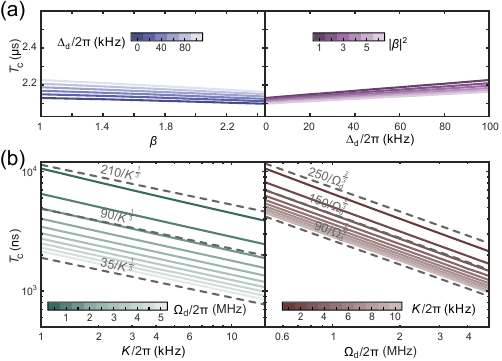}
  \caption{Numerically simulated evolutionary period of the quantum dynamics as a function of the coherent amplitude $\beta$, frequency detuning $\Delta_\mathrm{d}$ \textbf{(a)}, and Kerr nonlinearity $K$, drive strength $\Omega_\mathrm{d}$ \textbf{(b)}. Dashed lines serve as reference for the $T_\mathrm{c}$ versus $K^{-1/3}$ ($\Omega_\mathrm{d}^{-2/3}$) relationship on a logarithmic scale.}
  \label{cycleT2D_sim}
\end{figure}

\subsubsection{Numerical simulation}

In addition, we also perform systematic numerical simulations to investigate the relationship between the evolutionary period $T_\mathrm{c}$ and parameters $K$, $|\beta|^2$, $\Omega_\mathrm{d}$, and $\Delta_\mathrm{d}$. This is achieved by solving the master equation with varying these parameters across a broad parameter range. The simulated results are presented in Supplementary Figure~\ref{cycleT_sim}.

 As shown in Supplementary Figure~\ref{cycleT_sim}(a), the coherent state amplitude $\beta$ and the frequency detuning $\Delta_\mathrm{d}$ have little impact on the period of the cyclic dynamical evolution. In contrast, the Kerr nonlinearity $K$ and the drive strength $\Omega_\mathrm{d}$ can significantly alter the evolution period. This relationship is illustrated in Supplementary Figure~\ref{cycleT2D_sim}(b), where a linear fit on a logarithmic-logarithmic scale indicates a power-law dependence of the period $T_\mathrm{c}$ on $K$ (with an exponent of approximately $-1/3$) and on $\Omega_\mathrm{d}$ (with an exponent of approximately $-2/3$).

\begin{figure}[tb]
  \centering
  \includegraphics{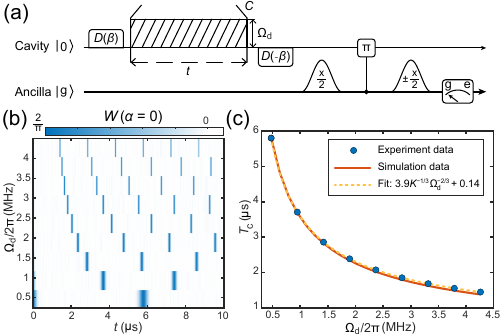}
  \caption{Experimental characterization of the cyclic evolution of the driven Kerr nonlinear oscillator.
\textbf{(a)} Experimental sequence for demonstrating the cyclic evolutionary dynamics of the detuned driven Kerr Hamiltonian.
  \textbf{(b)} Measured Wigner functions of the cavity state at the origin point in phase space as a function of the evolution time $t$ and the drive amplitude $\Omega_\mathrm{d}$.
  \textbf{(c)} Quantum evolution period extracted from (b) as a function of the drive strength $\Omega_\mathrm{d}$, exhibiting good agreement with the simulation results (solid line). The yellow dashed line is a fitting of the expression $T_\mathrm{c}=aK^{-1/3}\Omega_\mathrm{d}^{-2/3}+b$ with fitting parameters $a=3.9$ and $b=0.14$.}

  \label{cycle_exp}
\end{figure}

\subsubsection{Experimental demonstration}
The cyclic evolutionary dynamics are also demonstrated experimentally, with the experimental sequence shown in Supplementary Figure~\ref{cycle_exp}(a). The quantum dynamics of the Wigner function at the origin point in phase space are experimentally measured as a function of the drive strength $\Omega_\mathrm{d}$. The experimental results shown in Supplementary Figure~\ref{cycle_exp}(b) indicate that the cyclic period of the quantum state evolution decreases as the drive strength $\Omega_\mathrm{d}$ increases. We extract the evolutionary period $T_\mathrm{c}$ as a function of $\Omega_\mathrm{d}$, as shown in Supplementary Figure~\ref{cycle_exp}(c). The experimental results agree well with numerical simulations and fit well with the approximate relationship of $T_\mathrm{c}\propto K^{-1/3}\Omega_\mathrm{d}^{-2/3}$, indicating the successful modeling of the quantum dynamics of the detuned driven Kerr Hamiltonian.

\subsection{Influence of single-photon loss}
Additionally, the demonstrated quantum squeezing amplification method is also applicable in scenarios where the Kerr nonlinearity $K$ is extremely weak, even smaller than the single-photon decay rate $\kappa_\mathrm{c}$ of the storage cavity mode. To clarify this point, we conduct a numerical simulation that accounts for photon loss effects by solving the master equation:
\begin{eqnarray}
\frac{\mathrm{d}\rho}{\mathrm{d}t}=-i[H_\mathrm{d},\rho]+\kappa_\mathrm{c}\mathcal{D}[a]\rho,
\label{me}
\end{eqnarray}
where $\mathcal{D}[a]\rho=a\rho a^\dagger-1/2(a^\dagger a\rho+\rho a^\dagger a)$ is the Lindblad dissipative superoperator describing single-photon loss at a rate of $\kappa_\mathrm{c}$. In the simulation, we vary the ratio of $\kappa_\mathrm{c}/K$ and calculate the squeezing parameters $|\xi|$ as a function of the number of evolution cycles while adjusting the driving parameters to ensure similar squeezing parameters after the first evolution cycle for $\beta=2$. The simulation results, shown in Supplementary Figure~\ref{decay}, clearly reveal that the squeezing parameter increases with the number of cycles for different ratios of $\kappa_\mathrm{c}/K$, demonstrating the effectiveness of the squeezing amplification scheme with an extremely weak Kerr nonlinearity.

\begin{figure}[tb]
  \centering
  \includegraphics{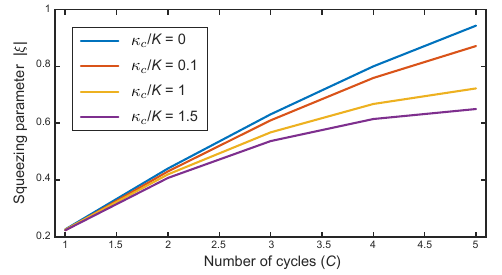}
  \caption{Numerical simulated squeezing parameters as a function of the evolution cycles for various ratios of $\kappa_\mathrm{c}/K$.}
  \label{decay}
\end{figure}

\begin{figure}[b]
  \centering
  \includegraphics{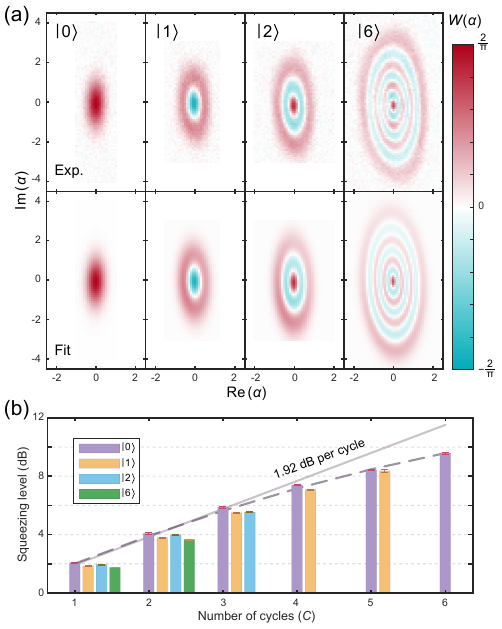}
  \caption{Generation and characterization of squeezed Fock states with the driven squeezing approach. 
  \textbf{(a)} The measured Wigner functions (up row) and their corresponding $\mathrm{2D}$ fitting results (bottom row) for the generated squeezed Fock states $|N\rangle$ (with $N=0,1,2,$ and $6$) after the second cycle $C=2$.
  \textbf{(b)} The extracted squeezing levels obtained from 2D fits to the measured Wigner functions of the generated squeezed Fock states as a function of the number of cycles. Error bars are the estimated 95\% confidence intervals of the fittings.  The solid line is a linear fit to the squeezing level for vacuum states in the linear region, showing an average addition in squeezing level of $1.92$~dB per cycle. The dashed line represents squeezing levels for vacuum states from numerical simulation.}
  \label{sr_fock}
\end{figure}

\begin{nonumsection}
    \section{Supplementary Note \thesection~-~Calculating the squeezing parameters}\label{sqzparam}
\end{nonumsection}

The generated squeezed states at each evolution cycle are characterized by measuring their Wigner functions and comparing them to those of ideal squeezed states. A general squeezed Fock state can be expressed as
\begin{eqnarray}
|\xi, N\rangle=S(\xi)|N\rangle,
\end{eqnarray}
where $S(\xi)=\exp{(\frac{\xi^*}{2}a^2 - \frac{\xi}{2}{a^\dagger}^2 )}$ represents the squeezing operator, $\xi=|\xi|e^{i\varphi}$ is the squeezing parameter with amplitude $|\xi|$ and phase $\varphi$, and $|N\rangle$ is the multiphoton Fock state with photon number $N$. The Wigner function~\cite{Al-Kader2003} of this state can be written as 
\begin{eqnarray}
W(\alpha)=\frac{2}{\pi}(-1)^Ne^{-2|\nu|^2}\mathcal{L}_N(4|\nu|^2),
\label{W}
\end{eqnarray}
where $\nu=\cosh(|\xi|)\alpha^*+e^{-i\varphi}\sinh(|\xi|)\alpha$, and $\mathcal{L}_N$ denotes the $N$-th order Laguerre polynomial function. We use this function to perform a two-dimensional (2D) fit to the measured Wigner functions of the generated squeezed multiphoton Fock states. In Supplementary Figure~\ref{sr_fock}(a), we present the 2D fitting results for the measured Wigner functions of the generated squeezed Fock states $|N\rangle$ (with $N=0,1,2,$ and $6$) after the second cycle $(C=2)$. The fitted results agree well with the experimentally measured Wigner functions. The squeezing parameters $\xi$ of the generated squeezed Fock states are finally extracted from the fitting results, as presented in Supplementary Figure~\ref{sr_fock}(b).

\begin{figure}[b]
  \centering
  \includegraphics{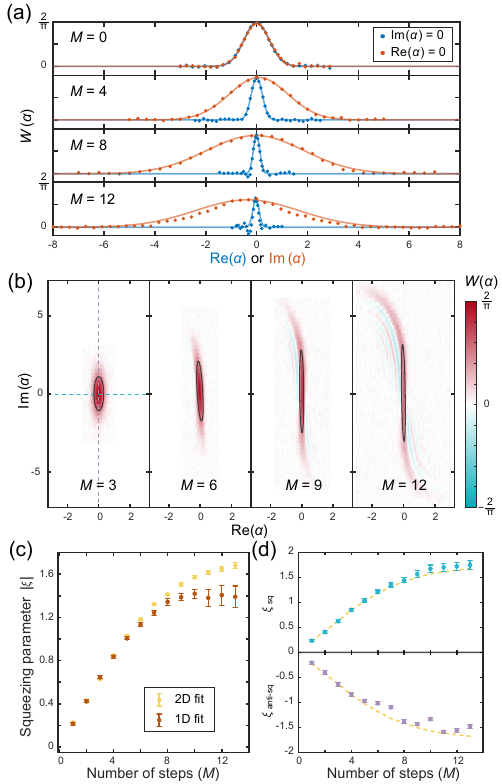}
  \caption{Measuring squeezing parameters with both 1D and 2D Wigner function fitting methods.
  \textbf{(a)} Measured 1D Wigner function cuts along the real axis (blue symbols) and the imaginary axis (red symbols) for different Trotter steps with $\beta=10$. Solid lines are global fits to the two traces. 
  \textbf{(b)} Measured 2D Wigner functions for different Trotter steps with $\beta=10$. The black elliptic lines represent the 2D fitting results.
  \textbf{(c)} Extracted squeezing parameters as a function of the number of Trotter steps with both the 1D and 2D Wigner function fitting methods shown in (a-b).
  \textbf{(d)} Extracted squeezing and anti-squeezing parameters from individually fitting the orthogonal quadrature cuts of the measured Wigner functions in (b). Dashed lines are the extracted squeezing parameters from 2D fit in (c). Error bars are the estimated $95\%$ confidence intervals of the fittings.}
  \label{sr}
\end{figure}

However, this 2D Wigner function fitting method requires measuring the full Wigner function of the cavity state to ensure a reliable extraction of the squeezing parameter $\xi$, and thus is time-consuming in the experiment. Here, we propose to perform a global fit on the one-dimensional (1D) Wigner function cuts along two orthogonal quadratures for the generated squeezed states. In our experiment, we first eliminate the rotation phase $\varphi$ of the squeezed state according to the method described in Supplementary Note~\ref{rotation}, and then measure the 1D Wigner function cuts along the real axis ($\mathrm{Im}(\alpha)=0$) and the imaginary axis ($\mathrm{Re}(\alpha)=0$), respectively. The ideal expressions for these two 1D Wigner function cuts can be derived as $W_{\mathrm{sq}}(\alpha)=\frac{2}{\pi} \exp{\left(-2e^{2|\xi|}\alpha^2\right)}$ for the squeezing quadrature and $W_{\mathrm{anti-sq}}(\alpha)=\frac{2}{\pi} \exp{\left(-2e^{-2|\xi|}\alpha^2\right)}$ for the anti-squeezing quadrature from Supplementary Equation~(\ref{W}). According to these two Gaussian functions, we perform a global fit to the two measured 1D Wigner function traces to extract the squeezing parameters $|\xi|$. Supplementary Figure~\ref{sr}(a) presents the 1D fitting results of the squeezed states generated using the Trotterization scheme (discussed in Supplementary Note~\ref{Trotter}) with $\beta=10$.

In addition, we directly measure the full 2D Wigner functions of these generated squeezed states and perform a 2D Wigner function fit, with the fitting results indicated by black elliptical lines in Supplementary Figure~\ref{sr}(b). The squeezing parameters extracted from both 1D and 2D Wigner function fitting methods are compared in Supplementary Figure~\ref{sr}(c), yielding consistent results for small squeezing. However, for large squeezed states, the fitting results have a little discrepancy because the generated squeezed states exhibit slight deformation and rotation in phase space due to the remaining Kerr and detuning Hamiltonian terms. This leads to fitting errors in the squeezing parameters when using the 1D Wigner function fitting method. The 2D Wigner function fit accounts for residual rotation phases in the fit, thus providing more accurate fitting results for the extracted squeezing parameters. 

To separately quantify the squeezing and anti-squeezing levels, we also individually fit the two orthogonal quadrature cuts of the measured Wigner function in Supplementary Figure~\ref{sr}(b). The individually extracted squeezing and anti-squeezing parameters as a function of Trotter steps are presented in Supplementary Figure~\ref{sr}(d), and are consistent with those obtained from the global 2D Wigner fits, validating the effectiveness and reliability of the fitting methodology.

\begin{nonumsection}
    \section{Supplementary Note \thesection~-~Quantum squeezing amplification with Trotterization technique}\label{Trotter}
\end{nonumsection}

\subsection{First- and second-order Trotterization}
The quantum evolutionary dynamics of a coherent state $|\beta\rangle$ under the driven Kerr Hamiltonian $H_\mathrm{d}$ in Supplementary Equation~(\ref{Hd}) can be understood by performing a displacement transformation, resulting in a Hamiltonian:
\begin{eqnarray}
H_{\beta}&=&D(-\beta)H_\mathrm{d}D(\beta)\notag\\
&=&\Delta'a^\dagger a-\frac{K}{2}{a^\dagger}^2a^2-\frac{K}{2}(\beta^2{a^\dagger}^2+\beta^{*2}{a}^2)\notag\\
&-&K\beta a^\dagger(a^\dagger a-r)+ \mathrm{H.C.}, \label{Hbeta}
\end{eqnarray}
with $\Delta'=\Delta_\mathrm{d}-2K|\beta|^2$ and $r=\Delta_\mathrm{d}/K + \Omega_\mathrm{d}/K\beta - |\beta|^2$ representing the photon-blockade parameter. To completely eliminate the undesired photon-blockade term, we add the driven Kerr Hamiltonian in two opposite displacement frames, $H_\beta$ and $H_{-\beta}$, resulting in the well-known Kerr parametric oscillator Hamiltonian:
\begin{eqnarray}
H_\mathrm{KPO} &=& \frac{H_\beta + H_{-\beta}}{2}\notag\\
               &=& -\frac{K}{2}(\beta^2{a^\dagger}^2+\beta^{*2}{a}^2)\notag\\
               &+&\Delta' a^\dagger a-\frac{K}{2}{a^\dagger}^2a^2.
\label{HKPO}
\end{eqnarray}
The time evolution of the Hamiltonian $H_\mathrm{KPO}$ can be approximated as
\begin{eqnarray}
e^{-iH_\mathrm{KPO}T} &=& (e^{-i\frac{H_{-\beta}+H_{\beta}}{2}\delta t})^M\notag\\
&=&(e^{-iH_{-\beta} \delta t/2} e^{-iH_{\beta}\delta t/2})^M \notag\\
&+& \mathcal{O}(1/{M}),
\label{trotter_1st}
\end{eqnarray}
using the Trotterization technique. Here, $\delta t=T/M$ is the discretized evolutionary time step, and $M$ is the number of Trotter steps. The Trotter errors in the first-order Trotter formula of Supplementary Equation~(\ref{trotter_1st}) can be suppressed to an order of $\mathcal{O}(1/{M})$~\cite{Suzuki1976,Han2021}, implying that a smaller time step would reduce the Trotter error. 

In our experiment, we calibrate the Trotter time step $\delta t$ by investigating the performance of quantum squeezing generation using the experimental sequence depicted in Supplementary Figure~\ref{dt}(a). In this sequence, the frame transformation displacement operation $D(\beta)$ has a short pulse duration of 10~ns to suppress Kerr evolution during the frame transformation. For a fixed total evolution time of $960$~ns, the measured 1D Wigner function cuts along the real axis ($\mathrm{Im}(\alpha)=0$) and the imaginary axis ($\mathrm{Re}(\alpha)=0$) as a function of the Trotter time step $\delta t$ are shown in Supplementary Figure~\ref{dt}(b). The experimental results clearly indicate that a smaller $\delta t$ would reduce the squeezing degree, while a larger $\delta t$ would increase the Trotter error. In our experiment, we choose an optimal Trotter time step of $\delta t=80$~ns as a trade-off.

\begin{figure}[tb]
  \centering
  \includegraphics{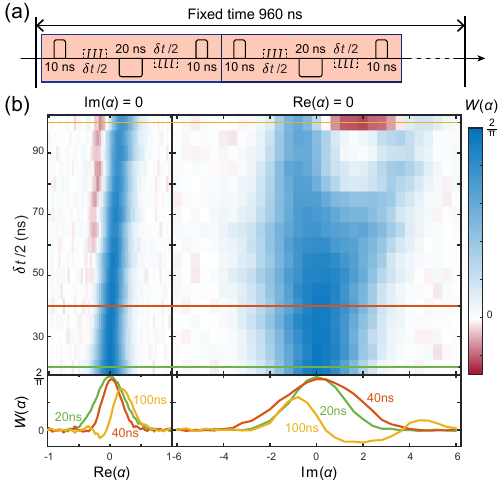}
  \caption{Calibrating the Trotter time step $\delta t$. 
  \textbf{(a)} Experimental sequence with a fixed total evolution time of $960$ ns.
  \textbf{(b)} Measured 1D Wigner function cuts along the real and imaginary axes for various Trotter time steps $\delta t$ with $\beta=8$. Three horizontal cuts are provided in the bottom row, with $\delta t/2=40$~ns (marked by red lines) being used in our experiment in the main text. }
  \label{dt}
\end{figure}

\begin{table}[t]
\caption{Optimal virtual phase shift $\phi_\mathrm{d}$} for various $|\beta|^2$ with the Trotterization method.
\begin{tabular}{p{2cm}<{\centering} p{4cm}<{\centering}}
  \hline
  \hline
  $|\beta|^2$ & $\phi_\mathrm{d}$ (rad) \\\hline
  &\\[-0.9em]
  1  & 4.8$\times10^{-4}\pi$ \\
  &\\[-0.9em]
  4 & 4.0$\times10^{-3}\pi$\\
  &\\[-0.9em]
  16 & 1.6$\times10^{-2}\pi$\\
  &\\[-0.9em]
  36 & 2.8$\times10^{-2}\pi$\\
  &\\[-0.9em]
  64 & 4.5$\times10^{-2}\pi$\\
  &\\[-0.9em]
  100 & 5.7$\times10^{-2}\pi$\\
  &\\[-0.9em]
  \hline
  \hline
\end{tabular}
\label{TableS2}
\end{table}

To completely eliminate the photon-blockade Hamiltonian term with the Trotterization method, the detuned drive amplitude $\Omega_\mathrm{d}$ in $H_{\beta}$ should be changed to $-\Omega_\mathrm{d}$ in $H_{-\beta}$ for each Trotter step. In our experiment, we set $\Omega_\mathrm{d}=0$ to limit the average photon number of intermediate states during the evolution and engineer a virtual phase shift $\phi_\mathrm{d}$ after each evolution cycle to partially counteract the accumulated phase induced by the remaining Kerr term. The optimal virtual phase shifts $\phi_\mathrm{d}$ for achieving the maximum squeezing parameter for various displacement amplitudes $|\beta|^2$ are determined from numerical simulations and are provided in Supplementary Table~\ref{TableS2}.

\begin{figure}[b]
  \centering
  \includegraphics{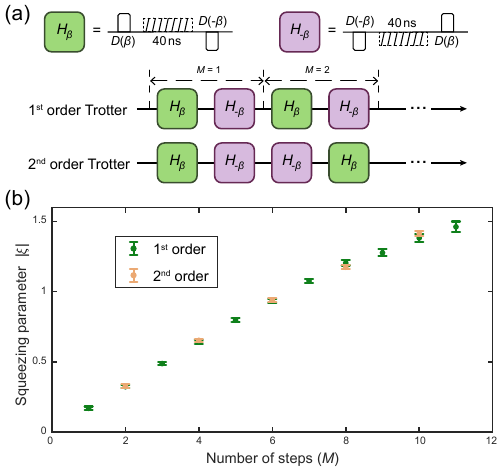}
  \caption{Comparison of the first- and the second-order Trotter schemes. Experimental sequences \textbf{(a)} and the measurement results \textbf{(b)} for both the first- and the second-order Trotter schemes with $\beta=8$. Error bars are the estimated $95\%$ confidence intervals of the $\mathrm{1D}$ fittings.}
  \label{trotter_2nd}
\end{figure}

In addition, Trotter errors can be further reduced by utilizing the second-order Trotter formula:
\begin{eqnarray}
    e^{-iH_\mathrm{KPO}T} &=& (e^{-i\frac{H_{-\beta}+H_{\beta}}{2}\delta t}e^{-i\frac{H_{-\beta}+H_{\beta}}{2}\delta t})^{\frac{M}{2}}\notag\\
    &=&(e^{-iH_{\beta} \delta t/2} e^{-iH_{-\beta}\delta t/2}e^{-iH_{-\beta}\delta t/2}e^{-iH_{\beta}\delta t/2})^{\frac{M}{2}} \notag\\
    &+& \mathcal{O}(1/{M^2})
\end{eqnarray}
where Trotter errors are suppressed to the order of $\mathcal{O}(1/{M^2})$ through rearrangement sequence of $H_{\beta}$ and $H_{-\beta}$, as illustrated in Supplementary Figure~\ref{trotter_2nd}(a). We directly perform an experimental comparison of the first- and second-order Trotter decomposition schemes and measure the squeezing parameters of generated squeezed vacuum states. The experimental results, shown in Supplementary Figure~\ref{trotter_2nd}(b), indicate almost identical performance for both the first- and second-order Trotter schemes. Therefore, it is unnecessary to adopt a higher-order Trotter scheme~\cite{Suzuki1993} in our experiment, and all experimental results presented in the main text are obtained using the first-order Trotter scheme.

\subsection{Squeezing phase calibration}\label{rotation}

During the Kerr evolution for generating squeezed state, a rotation phase $\varphi$ will accumulate due to the remaining Kerr and detuning Hamiltonian terms, resulting in a phase-space inclination of the squeezed states, as illustrated in Supplementary Figure~\ref{scanPhase}(a). It is crucial to characterize and eliminate this phase to obtain an accurate squeezing parameter from the 1D Wigner function fit discussed in Supplementary Note~\ref{sqzparam}. In our experiment, this phase is calibrated by measuring the Wigner function $W(\alpha)$ of the generated squeezed state as a function of the rotation phase $\mathrm{arg}(\alpha)$ with a fixed amplitude $|\alpha|=0.7$. The experimental results are shown in Supplementary Figure~\ref{scanPhase}(b) for the generated squeezed states at each Trotter step with $\beta=10$. Subsequently, the rotation phase of the squeezed state at each Trotter step is extracted by identifying the rotation angle that maximizes the Wigner functions. Finally, the calibrated phases of the squeezed states are eliminated by performing virtual phase rotations in the experimental sequence. These virtual operations are implemented by adding a phase offset to the subsequent microwave pulses in the sequence and have a duration of zero~\cite{mckay2017}.

\begin{figure}[h]
  \centering
  \includegraphics{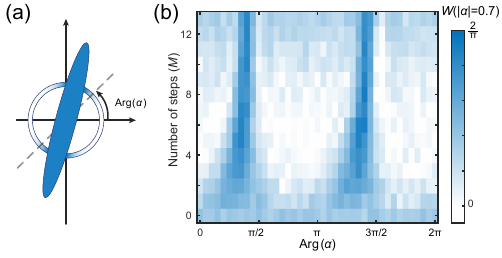}
  \caption{Calibrating the rotation phases of generated squeezed states.
  \textbf{(a)} Schematic illustration for calibrating the rotation phases of generated squeezed states. The ring structure indicates the circumferential 1D Wigner functions.
  \textbf{(b)} Experimentally measured 1D Wigner functions $W(\alpha)$ as a function of the rotation phase $\mathrm{arg}(\alpha)$ with a fixed amplitude $|\alpha|=0.7$ for various Trotter steps $M$ with $\beta=10$.
  }
  \label{scanPhase}
\end{figure}

\begin{nonumsection}
    \section{Supplementary Note \thesection~-~Quantum metrology for sensing displacement}
\end{nonumsection}
\subsection{Parity measurement}
A variety of protocols for quantum-enhanced precision measurement of small displacements have been well established, such as homodyne detection~\cite{Fadel2024}, projective measurement~\cite{Pan2025,Park2025}, and parity measurement~\cite{deng2024}. In this work, we employ the parity measurement for high-precision displacement estimation to evaluate the metrological capability of the generated squeezed states. This is achieved by first encoding a displacement amplitude $\alpha$ onto the generated squeezed state via a displacement operation, and then probing it through an ancilla-assisted parity measurement, as shown in Fig.~4a of the main text. The displacement amplitude can be inferred from the qubit ground state population, which relates to the Wigner function of an ideal squeezed state through the formula $P_g^{\mathrm{ideal}}(\alpha)=\frac{1}{2}[\frac{\pi}{2}W_{\mathrm{sq}}(\alpha)+1]=\frac{1}{2}[\exp{\left(-2e^{2|\xi|}\alpha^2\right)}+1] $. We fit the measured results with the probability model $P_g(\alpha)=A\exp(-2e^{2|B|}\alpha^2)+D$ to obtain a smooth probability distribution $P_g(\alpha)$, where the free parameters $A$ and $D$ account for experimental imperfections, and $B$ is the extracted squeezing parameter. The corresponding classical Fisher information (FI) for this measurement strategy can thus be calculated by $I_\mathrm{F}(\alpha)=\frac{1}{P_g(1-P_g)}(\frac{\mathrm{d}P_g}{\mathrm{d}\alpha})^2$~\cite{PARIS2009}, as shown in Supplementary Figure~\ref{parity_results}(a). This metrology scheme exhibits an optimal operating point where the sensitivity to small displacement is maximized, that is $\alpha_{\mathrm{opt}}=\arg\max_\alpha I_\mathrm{F}(\alpha)$, as presented in Supplementary Figure~\ref{parity_results}(b). All experimental data presented in Fig.~4 of the main text and Supplementary Figure~\ref{deviation} are obtained at this optimal displacement amplitude. 

\begin{figure}[t]
  \centering
  \includegraphics{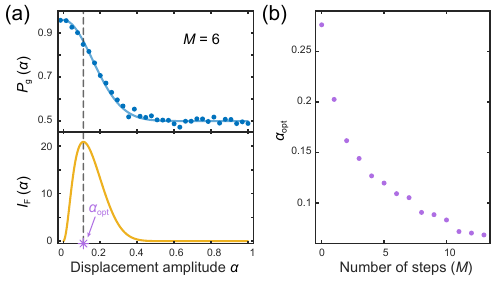}
  \caption{Extracting the optimal displacement amplitude for quantum sensing using the Trotterization-generated squeezed states.
  \textbf{(a)} Measured qubit ground state populations (blue symbols), corresponding fit (blue line), and extracted FI (yellow line) as a function of the displacement amplitude $\alpha$ with $M=6$. The optimal displacement amplitude $\alpha_\mathrm{opt}$ with maximum FI is indicated with a snowflake symbol.
  \textbf{(b)} Extracted optimal displacements as a function of the number of Trotter steps $M$.
  }
  \label{parity_results}
\end{figure}

\subsection{Bayesian inference}

Bayesian inference is a versatile and powerful methodology for parameter estimation based on probability distributions. In this work, we employ Bayesian estimation to extract the encoded small displacement amplitude $\alpha$ from the measured qubit ground population.

To estimate a displacement parameter $\alpha$ given its prior probability $P(\alpha)$, the core of the Bayesian inference lies in maximizing the posterior probability 
\begin{eqnarray}
P(\alpha|\mathcal{M})\propto P(\alpha)\times P(\mathcal{M}|\alpha),
\label{posterior_P}
\end{eqnarray}
where $P(\mathcal{M}|\alpha)$ represents the likelihood of observing the ancilla in ground state ($\mathcal{M}=g$) or excited state ($\mathcal{M}=e$) under a small displacement $\alpha$. Assuming a uniform prior probability $P(\alpha)$ over the displacement amplitude interval $\alpha\in[0,0.5]$, Bayesian estimation reduces to maximum likelihood estimation (MLE). In this case, we have $P(\alpha|\mathcal{M})\propto P(\mathcal{M}|\alpha)$. In a series of $N_{\mathrm{meas}}$ independent experiments, the ancillary qubit is measured in the ground (excited) state in $N_g$ ($N_e$) instances. The corresponding likelihood function is given by $P(\mathcal{M}|\alpha)^{N_{meas}}=[P_g(\alpha)]^{N_g}\times[1-P_g(\alpha)]^{N_e}$. In our experiment, we employ the log-likelihood function,
\begin{eqnarray}
L(\mathcal{M}_1,\dots\mathcal{M}_{N_\mathrm{meas}}|\alpha)&=&P_g^{\mathrm{exp}}\ln(P_g(\alpha))\notag\\&+&P_e^{\mathrm{exp}}\ln(1-P_g(\alpha)),
\label{likelihood}
\end{eqnarray}
where $P_g^{\mathrm{exp}}=N_g/N_{\mathrm{meas}}$ and $P_e^{\mathrm{exp}}=N_e/N_{\mathrm{meas}}$ are the corresponding outcome probabilities. And the displacement parameter is finally estimated by maximizing the log-likelihood function through $\alpha_{\mathrm{est}}=\arg\max_\alpha L(\mathcal{M}_1,\dots\mathcal{M}_{N_\mathrm{meas}}|\alpha)$. The Bayesian posterior distribution $P(\alpha|\mathcal{M})$ directly allows extracting the variance of the estimated displacement, with the extracted standard deviation shown in Supplementary Figure~\ref{deviation}. To better quantify the correlated noise in the system, we utilize a well-established sensitive metric---Allan deviation~\cite{Allan1987}---to identify and quantify the low-frequency fluctuations, with the experimental results shown in Fig.~4b in the main text.

\begin{figure}[h]
  \centering
  \includegraphics{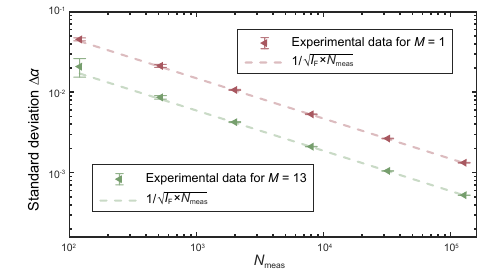}
  \caption{Extracted standard deviation (triangles) from the Bayesian posterior distributions for estimating the displacement amplitude using squeezed states generated with different Trotter steps (red: $M=1$, green: $M=13$) versus the number of independent measurements $N_\mathrm{meas}$. Dashed lines represent the theoretical Cram\'er-Rao bound using classical FI. Error bars are standard deviations from repeated experiments.
  }
  \label{deviation}
\end{figure}

\begin{nonumsection}
    \section{Supplementary Note \thesection~-~Breakdown of the dispersive approximation}\label{breakdown}
\end{nonumsection}

In pursuit of a high squeezing rate with the Trotterization technique, a substantial displacement amplitude is required during the displacement frame transformation. However, this leads to excessive photon numbers in the storage cavity, potentially disrupting the dispersive approximation shown in Supplementary Equation~(\ref{H_total}). Consequently, the Jaynes-Cummings (JC) interaction between the storage cavity and the ancilla qubit should be considered, as described by the Hamiltonian:
\begin{eqnarray}
H_\mathrm{JC} &=& \omega_\mathrm{c}' a^\dagger a +
\omega_\mathrm{q}' q^\dagger q -\frac{\eta_\mathrm{q}}{2}{q^\dagger}^2q^2\notag\\
&+&g_\mathrm{qc}(a^\dagger q+aq^\dagger),
\label{H_jcl}
\end{eqnarray}
where $\omega_\mathrm{c}'$ and $\omega_\mathrm{q}'$ are their respective bare
resonance frequencies and $g_\mathrm{qc}$ is the coupling constant between the
qubit and the storage cavity. This interaction can induce qubit excitations when the storage cavity contains a large number of photons. This phenomenon is also verified in our experiment by measuring the qubit excited state populations as a number of Trotter steps for various displacement amplitudes $|\beta|^2$, with the results depicted in Supplementary Figure~\ref{pe}. As the displacement amplitude increases, the qubit excited state population progressively increases with the number of Trotter steps. The breakdown of the dispersive approximation imposes a constraint on the maximum achievable value of $\beta$ in our experiment. In addition, the qubit excited population induced by the breakdown of the dispersive approximation can diminish the state fidelity of the generated squeezed state because the entanglement between the qubit and the storage cavity during the evolution would reduce the purity of the quantum state in the storage cavity.

\begin{figure}
  \centering
  \includegraphics{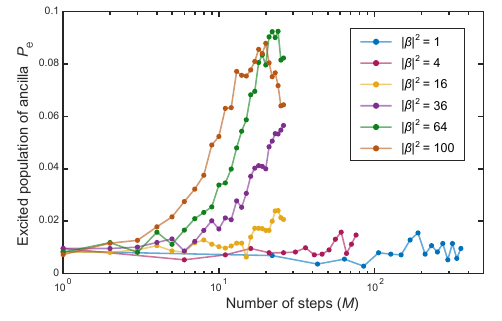}
  \caption{Excited population of the auxiliary qubit as a function of the number of Trotter steps for various displacement amplitudes $|\beta|^2$.
 }
  \label{pe}
\end{figure}

\begin{nonumsection}
    \section{Supplementary Note \thesection~-~Error analysis in the Trotter scheme}
\end{nonumsection}

\begin{figure} [b]
  \centering
  \includegraphics{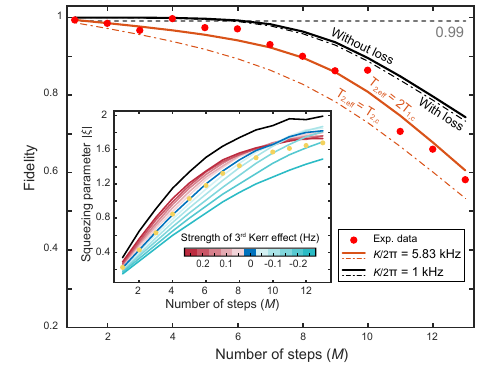}
  \caption{Error analysis of the generated squeezed states with Trotterization technique. Measured (red circles) and simulated (solid and dashed lines) state fidelities between the generated quantum states and ideal squeezed states as a function of the Trotter steps. The inset presents the squeezing parameters (yellow circles) extracted from 2D Wigner function fits shown in Supplementary Figure~\ref{sr}(c). Solid lines in the inset show the simulation results with varying the third-order Kerr nonlinearity in the simulation. Black lines represent the simulations with a Kerr nonlinearity of $K/2\pi=1$~kHz.}
 
  \label{error}
\end{figure}

The primary error sources for the displacement-enhanced squeezing generation with the Trotterization technique include decoherence errors of the storage cavity, Trotter errors, deformation errors caused by the Kerr nonlinear term, ancillary qubit excitation errors resulting from the breakdown of the dispersive approximation, and errors due to higher-order Kerr nonlinearity. In order to estimate the contributions of these errors, we conduct numerical simulations by solving the master equation while accounting for these error sources. Supplementary Figure~\ref{error} presents both the simulated and measured state fidelities of the generated squeezed states as a function of Trotter steps with $\beta=10$. The fidelities are determined by comparing the simulated or reconstructed density matrices to the ideal squeezed state. The Trotterized squeezing approach can benefit from the echoed displacements, which effectively suppress cavity dephasing errors during the evolution. This is confirmed by comparing the simulation results with and without including the cavity pure dephasing rate in the simulation, as presented in Supplementary Figure~\ref{error}. When we set an effective cavity decoherence lifetime $T_\mathrm{2,eff}=2T_\mathrm{1,c}$ (indicating no pure dephasing rate) in the master equation solver, the simulated fidelities are in good agreement with the measurements. This alignment confirms the effectiveness of error modeling and the suppression of cavity dephasing in our Trotterized squeezing generation method. Additionally, we find that the deformation errors induced by Kerr nonlinearity are the predominant errors, which increase with the number of Trotter steps. 

Besides, we also provide a comparative analysis of simulated versus experimentally measured squeezing parameters in the inset of Supplementary Figure~\ref{error} for the generated squeezed states across various Trotter steps. When considering only the first-order $(\frac{K}{2}a^{\dagger2}a^2)$ and second-order $(\frac{K_2}{6}a^{\dagger3}a^3)$ Kerr nonlinearities in the simulation, we find that the simulation curve of the squeezing parameters increasingly diverges from the experimental results as the Trotter steps increase. This divergence may be attributed to errors induced by higher-order Kerr nonlinearity, such as the third-order Kerr nonlinearity term $(\frac{K_3}{24}a^{\dagger4}a^4)$. Here, $K_{2}$ and $K_3$ denote the strengths of the second-order and third-order Kerr effects, respectively. Given that this extremely weak higher-order Kerr nonlinearity is beyond our experimental characterization precision, we investigate its influence by performing numerical simulations with varying the strengths of the third-order nonlinearity. The simulation results shown in the inset capture the behavior of the measured squeezing parameters over the Trotter steps.

\begin{nonumsection}
    \section{Supplementary Note \thesection~-~Squeezing approaches comparison}
\end{nonumsection}

In this section, we analyze and compare various methods for generating microwave squeezed states in superconducting quantum circuits. In this system, flying squeezed microwave states have been successfully generated using Josephson parametric amplifiers, achieving a maximum squeezing of 11.4 dB~\cite{yurke1988,qiu2023}. Besides, microwave states within the cavity in this system can also be compressed to realize two-photon squeezing operations. This can be achieved by employing a superconducting qubit through measurement and post-selection, resulting in a squeezing of 5.7 dB~\cite{wang2017}. These intracavity squeezed states can also be generated using a parametric mixing pump of the nonlinear Josephson elements~\cite{dassonneville2021,marti2024,eriksson2024}, achieving squeezed microwave phononic states beyond 3 dB~\cite{marti2024} and photonic states up to 8.63 dB squeezing~\cite{eriksson2024}. Furthermore, parameterized quantum circuits with echoed conditional displacements have been employed to generate squeezed states within the microwave cavity, reaching a squeezing up to 11.1 dB~\cite{eickbusch2022,pan2023}.

In comparison to these squeezing methods, our Trotterized squeezed approach generates squeezed states deterministically, utilizing simple and natively available gates and dynamics. These generated squeezed states within the cavity differ from itinerant squeezing due to the long coherence lifetime of the cavity mode, enabling efficient quantum information processing with these intracavity squeezing resources. Unlike other squeezing approaches, our method eliminates decoherence errors induced by the ancilla qubit during the evolution. Additionally, the echoed displacement operations effectively suppress cavity dephasing errors during the evolution. With these advantages, our squeezing approach achieves a maximum squeezing degree of 14.6 dB, which, to our knowledge, is the largest squeezing value for intracavity microwave photonic states.

%